%% file: main.tex
\documentclass[
    reprint,
    superscriptaddress,
    amsmath,
    amssymb,
    aps,
    pra,
    longbibliography
]{revtex4-2}

\usepackage{changes}

\setdeletedmarkup{\color{blue}\sout{#1}}

\usepackage[normalem]{ulem}% fix the bibliography format
\usepackage{amsthm}
\usepackage{float}

\usepackage{graphicx}% Include figure files
\usepackage{dcolumn}% Align table columns on the decimal point
\usepackage{bm}% bold math
\usepackage{hyperref}% add hypertext capabilities
\usepackage{ulem}
\usepackage{braket}
\usepackage{color}
\usepackage{xspace}
\usepackage{mhchem}
\usepackage{graphicx}
\usepackage{ulem}
\usepackage{url} 

\usepackage{float}
\makeatletter
\let\newfloat\newfloat@ltx
\makeatother

\usepackage{algorithm}
\usepackage{algpseudocode}

\input{tikzcommand}

\input{commands}

\newtheorem{theorem}{Theorem}

\begin{document}

\setlength{\parindent}{15pt}
\sloppy

\preprint{NoiseQC}

\title{
Harnessing Intrinsic Noise for Quantum Simulation of Open Quantum Systems
}

\author{Sameer Dambal}
\affiliation{
Theoretical Division, Los Alamos National Laboratory, Los Alamos, NM, 87544, USA.
}
\affiliation{Department of Physics, University of Houston, Houston, TX 77204, USA}

\author{Akira Sone}
\email{akira.sone@umb.edu}
\affiliation{Department of Physics, University of Massachusetts, Boston, MA 02125, USA}
\affiliation{The NSF AI Institute for Artificial Intelligence and Fundamental Interactions}

\author{Yu Zhang}
\email{zhy@lanl.gov}
\affiliation{
Theoretical Division, Los Alamos National Laboratory, Los Alamos, NM, 87544, USA.
}

\date{\today}

\begin{abstract}
Simulating open quantum systems on quantum computers is challenging because dissipative dynamics are nonunitary, whereas gate-based quantum algorithms are typically built from unitary operations. 
Conventional approaches can overcome this mismatch by embedding nonunitary processes into enlarged unitary circuits, often with ancillary-qubit and gate overhead.
In this work, we propose an alternative perspective in which we selectively preserve the physical noise to emulate nonunitary channels. More specifically, we present a partial encoding algorithm that iteratively leverages intrinsic noise to mimic the dynamics of a target open-system. We benchmark the convergence of this algorithm against the dissipative exciton transport Hamiltonian and retrieve their dynamics. By utilizing intrinsic hardware noise to emulate the target nonunitary dynamics, this approach avoids the dual overhead of error-correcting that noise and using ancillary qubits to re-introduce nonunitarity. This significantly reduces both qubit requirements and physical fidelity thresholds, paving the way for more efficient techniques to simulate open quantum systems on quantum computers.
\end{abstract}

\maketitle

\section{Introduction}
\label{introduction}
In realistic conditions, a quantum system inevitably interacts with its bath (or environment), thereby forming an \textit{open quantum system}. Consequently, the study of open quantum dynamics is essential for understanding the properties and behaviors of molecules, materials, and quantum devices under experimentally relevant conditions. Unlike static properties, dynamical simulations capture transient behavior~\cite{kashuba2013transient, giorgi2019transient, schulz2018numerical}, reveal kinetic effects~\cite{juzar2017kinetics, pryadko2006quantum}, and elucidate nonequilibrium processes such as transport~\cite{dhawan2024anomalous, elenewski2014real, mostame2012quantum} and quantum decoherence~\cite{nielsen2010quantum,gardiner2004quantum,breuer2002theory,lindblad1976generators,manzano2020short,zurek2003decoherence, dambal2025nt}. They also play a significant role in studying driven systems~\cite{dagvadorj2015nonequilibrium, carrega2016energy, diehl2008quantum, Chen:2018aa, Zhang:2013ab}, quantum information flow~\cite{lu2010quantum,hsieh2019nonmarkovianity,deffner2020quantum,deffnerquantum2013,santos2025quantifying}, and quantum technologies including sensors~\cite{degen2017quantum,giovannetti2006quantum,giovannetti2011advances,barbieri2022optical,zhou2020quantum,zhang2021distributed,sone2024conditional}, transducers~\cite{heugel2019quantum, zeuthen2020figures, zhuang2025quantum, wang2022quantum, hou2025correlated}, and simulators~\cite{georgescu2014quantum, altman2021quantum, bauer2023quantum, endo2020variational, omalley2016scalable, domanti2024floquet, feldmeier2024quantum, monroe2021programmable}. However, accurate simulation of open quantum dynamics remains notoriously challenging~\cite{li2020exponential, Ding:2024vg, biele2012stochastic, Schlimgen:2021vs}, since the Hilbert space required to represent the quantum state grows exponentially with system size and rapidly exceeds classical computational limits.

State-of-the-art algorithms on classical computers, such as Density Matrix Renormalization Group (DMRG)~\cite{schollwock2005density} and Multi-Configurational Time-Dependent Hartree (MCTDH)~\cite{beck2000multiconfiguration, meyer2009multidimensional}, tackle this challenge by truncating the Hilbert space via low-rank approximations~\cite{markovsky2012low}. While highly successful in low-dimensional or weakly entangled systems, these methods become less effective for high-dimensional, strongly correlated, or long-range interacting systems~\cite{baiardi2020density}. In addition, Hierarchical equations of motion (HEOM)~\cite{jin2008exact, tanimura2020numerically} offer an alternative route that avoids standard approximations such as the Born, Markovian, and rotating-wave approximations, but at substantial computational cost. In particular, at low temperatures, convergence of the Matsubara expansion requires many terms that dramatically increases the hierarchy depth and makes simulations extremely expensive~\cite{Hu:2011aa, Zhang:2013aa}. As a result, the simulation of many realistic open quantum dynamics problems often falls beyond classical reach.

Richard Feynman proposed in 1982 that quantum computers offer a natural platform for modeling quantum systems, including quantum dynamics~\cite{feynman2018simulating}. 
The number of amplitudes needed to describe a generic quantum state grows exponentially with the number of qubits, which motivates quantum simulation as a promising route for selected problems, although any quantum advantage remains problem- and algorithm-dependent.
This promise stimulated the development of quantum algorithms for energy estimation and quantum dynamics~\cite{Miessen:2023aa, Alexeev:2025aa, Zhang:2022aa, Tkachenko:2024aa, Berthusen:2024aa, Bauer:2020aa, Cerezo:2021aa, Delgado-Granados:2025aa}.
However, performing such simulations is not a trivial task, as quantum computers are extremely fragile processors and are affected by thermal, chemical, electromagnetic, and other fluctuations that arise from the bath they are placed in. These noise sources can drive the output state from the intended target, necessitating quantum error-correction (QEC) codes that assemble high-fidelity logical qubits from large numbers of physical qubits~\cite{Acharya:2025aa, Fowler:2012aa}.

On the other hand, quantum algorithms for open quantum system dynamics face a fundamental mismatch: (fault-tolerant) quantum computers are inherently unitary, while open quantum system dynamics are nonunitary.  One common strategy to bridge this gap is Hilbert-space dilation~\cite{Hu:2020vx, Yu:2025aa}, in which ancilla qubits emulate bath degrees of freedom and the enlarged system evolves unitarily. Other strategies explored in the literature include unitary decompositions~\cite{Schlimgen:2021vs}, analog quantum simulation~\cite{Kim:2022wu}, and pseudo-identity unitary transformations~\cite{Rose:2025aa}, among others. Yet these approaches often demand additional qubits and deep circuits, compounding hardware overhead and error susceptibility.

In this work, we ask a different question: rather than suppressing all noise and nonunitary effects to restore perfectly unitary logical evolution, can one instead harness the native nonunitary dynamics of noisy qubits (whether they are physical qubits or lower-fidelity logical qubits) to emulate the nonunitary evolution of open quantum systems?
We answer this by developing a new approach in which we take a target open quantum system's decoherence channels and partially encode them as unitary gates on a quantum circuit. In doing so, we allow noise to modify the effective channel generated by the quantum circuit and use this information to iteratively encode the residual system channels across subsequent time steps. By running the circuit over multiple shots, we average over multiple trajectories and generate a map that effectively emulates the nonunitary dynamics of a target open quantum system.
This strategy offers two key advantages:  (i) it relaxes the requirement for ultra-high-fidelity logical qubits, lowering the physical-qubit and gate overhead of QEC, i.e., only the error components that exceed or mismatch the desired dissipation need to be corrected; and (ii) it bypasses the need to embed nonunitary evolution into unitary dynamics. Moreover, we show that the intrinsic noise triggers the formation of clusters of partially encoded system channels. These clusters act as decoherence-free subspaces (DFSs)~\cite{lidar1998decoherence,lidar1999concatenating,kwiat2000experimental,bacon1999robustness,lidar2001decoherence,wang2013numerical,bacon2000universal,sone2025no} that can also be used as resources when performing quantum simulations. 

This perspective is conceptually related to several established approaches that exploit, rather than eliminate, environmental interactions. Passive quantum error-correction techniques~\cite{liu2024dissipative, shirol2026passive, marshman2018passive, de2006passive}, including DFSs~\cite{lidar1998decoherence, lidar2003decoherence, kwiat2000experimental} and noiseless subsystems, exploit symmetries in system-environment couplings to encode quantum information into sectors of Hilbert space that are intrinsically insensitive to specific classes of noise, thereby reducing the need for active error-correction cycles. Dissipative quantum computing~\cite{mi2024stable, lin2025dissipative} and reservoir engineering take a complementary viewpoint by deliberately designing system-bath interactions such that dissipation autonomously prepares, stabilizes, or processes quantum states. More broadly, noise-assisted quantum information processing has revealed that decoherence and environmental fluctuations can enhance functionality, ranging from transport phenomena in biological and condensed-matter systems to state preparation, sensing, and error-suppression protocols. Our framework extends these ideas to open-system simulation by treating calibrated intrinsic hardware noise as a computational primitive that can directly implement target Pauli decoherence channels when the target and hardware channels are suitably matched.

The rest of the paper is organized as follows. In Sec.~\ref{sec:theory}, we introduce the theoretical framework for open quantum dynamics and present our partial encoding algorithms for leveraging intrinsic hardware noise. In Sec.~\ref{sec:bound}, we formalize the method using the Choi--Jamio{\l}kowski isomorphism and derive a practical bound on the residual channel mismatch. In Sec.~\ref{sec:results}, we analyze the residual-convergence behavior of the encoding procedure in the numerical examples of Subsec.~\ref{subsec:convergence} and benchmark the method on coherent-dissipative energy transport in a multi-exciton model in Subsec.~\ref{subsec:benchmarking}. Finally, in Sec.~\ref{sec:summary}, we summarize our results and discuss their implications and future directions.

\section{Theory}\label{sec:theory}

The Hamiltonian of an open quantum system $S$ interacting with a bath $B$ can be expressed as
\begin{equation}
    \label{eq:hamiltonian}
    {H}_T = {H}_S + {H}_B + {H}_{SB},
\end{equation}
where ${H}_S$, ${H}_B$, and ${H}_{SB}$ denote the Hamiltonians of the system, the bath, and the system–bath interactions, respectively.
The dynamics of the open quantum system can be derived by tracing out the bath degrees of freedom (DOFs) from the unitary dynamics of the entire system ${H}_T$. One of the widely used formalisms to model Markovian dynamics is given by the quantum Lindblad master equation~\cite{nielsen2010quantum,gardiner2004quantum,breuer2002theory,lindblad1976generators,manzano2020short, Peng:2026aa} (with $\hbar=1$),

\begin{equation}
\label{Lindblad_equation}
\frac{\partial \rho}{\partial t} = i[\rho, H_S] +\sum_i \left(L_i\rho L_i^{\dagger}-
\frac{1}{2}L_i^{\dagger}L_i\rho-\frac{1}{2}\rho L_i^{\dagger}L_i\right).
\end{equation}
The first term on the right-hand side contributes to the unitary dynamics of the system, while the second term contributes to the nonunitary dynamics arising from the coupling of the system with the bath degrees of freedom.

Beyond Markovianity, the dynamics of open quantum systems are generally formulated in the so-called Kraus representation~\cite{nielsen2010quantum}
\begin{equation}
\label{Kraus_formalism}
\mathcal{E}(\rho) = \sum_i K_i \rho K_i^\dagger,
\end{equation}
where ${K}_i$ are the Kraus operators satisfying the completeness relation $\sum_i K_i^\dagger K_i = \mathbb{I}$ and form a completely positive trace-preserving (CPTP) map~\cite{nielsen2010quantum}. The Kraus formalism can be readily derived from the unitary dynamics of the entire system by tracing out the bath degrees of freedom.
Note that when $K_0 = \mathbb{I} - (iH_S + \frac{1}{2}\sum_i L_i^\dagger L_i)\Delta t$ and $K_i = \sqrt{\Delta t}L_i$, Eq.~\eqref{Kraus_formalism} reduces to the Lindblad formalism from Eq.~\eqref{Lindblad_equation} as $\Delta t\to 0$. One can also convert Eq.~\eqref{Kraus_formalism} into an effective Pauli noise model using Pauli twirling techniques~\cite{wallman2016noise}. Averaging over randomized Pauli conjugations suppresses off-diagonal components in the Pauli representation of the channel, yielding a structured stochastic Pauli channel rather than eliminating the underlying physical noise. As a result, we describe the dissipation channels of a system with
\begin{equation}
\label{system_pauli_channel}
\mathcal{E}_{\text{sys}}(\rho) = \sum_i w_i {P}_i \rho {P}^\dagger_i,
\end{equation}
where ${P}_i \in \{{I}, {X}, {Y}, {Z}\}^{\otimes n}$ is an $n$-qubit Pauli string, and $\sum_i w_i = 1$.
%We observe that Eq.~\eqref{system_pauli_channel} converts a pure state into a mixed state and agrees well with the dynamics of open quantum systems. This behavior is also replicated in current noisy intermediate-scale quantum (NISQ) devices, where noise converts any pure input state into highly mixed output states. 
Similarly, the (calibrated) intrinsic hardware-noise channel of a quantum computer, with an architecture- and gate-dependent coupling map, is modeled as
\begin{equation}
\label{qc_channels}
\mathcal{E}_I(\rho) = \sum_k w_k {P}_k \rho {P}^\dagger_k.
\end{equation}

%For notational simplicity, Eq.~\eqref{qc_channels} suppresses possible dependence on simulation step, gate type, and hardware location. If the calibrated noise varies during a time-discretized simulation, we write the channel at step $t$ and for gate/location label $g$ as
%\[
% \mathcal{E}_I^{(t,g)}(\rho)=\sum_k w_k^{(t,g)} P_k\rho P_k^\dagger.
%\]
%The target weights may likewise be time dependent for a time-dependent CPTP target map. The partial-encoding update is unchanged: at each step the residuals are computed using the target weights and the currently calibrated intrinsic-noise weights. Thus, non-constant but calibrated stochastic Pauli noise can be incorporated step by step within the same framework.

%This holds true since the intrinsic noise channels are inherently the decoherence channels of a quantum computer that is coupled to its environment. 
To simulate an open-system described by Eq.~\eqref{system_pauli_channel}, we encode the system degrees of freedom onto qubits and implement the coherent dynamics as gates on the quantum computer. The target system channel and the intrinsic hardware-noise channel are generally different: $\{{P}_i\}_{\text{sys}} \neq \{{P}_k\}_{I}$ and $\{w_i\}_{\text{sys}} \neq \{w_k\}_{I}$. We therefore develop a protocol that partially encodes target system channels $\{{P}_i\}_{\text{sys}}$ on a quantum circuit and uses calibrated intrinsic noise $\{{P}_k\}_{I}$ to implement part of the residual target channel. We assume that a calibrated single- or two-qubit intrinsic noise channel is activated when the corresponding gate is implemented. The strength of the intrinsic noise sets a lower limit on this process and can cause certain target channels to be overencoded after several iterations. In such cases, selective mitigation or partial error correction is required only for the mismatched components, rather than for all intrinsic noise components. We now describe the partial encoding algorithm by first outlining the protocol for single-qubit channels and then extending it to two- and multi-qubit channels.

\begin{figure*}[!htb]
   \centering
   \input{figures/1D_braiding}
   \caption{a)~1D braiding structure with $d_c=d_b+1$.
   b)~3D braiding with the ${Y}_1{I}_2$ node. This braiding generates a closed channel cluster, analogous to a DFS in channel space, for the exemplified $\{{Y}_1{I}_2, {I}_1{Y}_2, {X}_1{Z}_2, {Z}_1{X}_2\}$ system channels under the $\{{X}_1{X}_2,{Y}_1{Y}_2,{Z}_1{Z}_2\}$ channels. Blue arrows indicate mapped channels; yellow arrows indicate alternate choices that preserve the closed cluster while maintaining the $d_c=d_b+1$ structure.
   c) The simplest example of a braiding structure with $d_c>d_b+1$. Here $d_c=4$ and $d_b=2$. This does not generate an all-to-all mapping because the absence of ${Y}_1{Y}_2$ means that one-step braiding does not close the full channel cluster. The black arrows in each subfigure denote the branches emerging out of a selected node and highlight the braiding dimension.}
   \label{fig:braiding-combined}
\end{figure*}
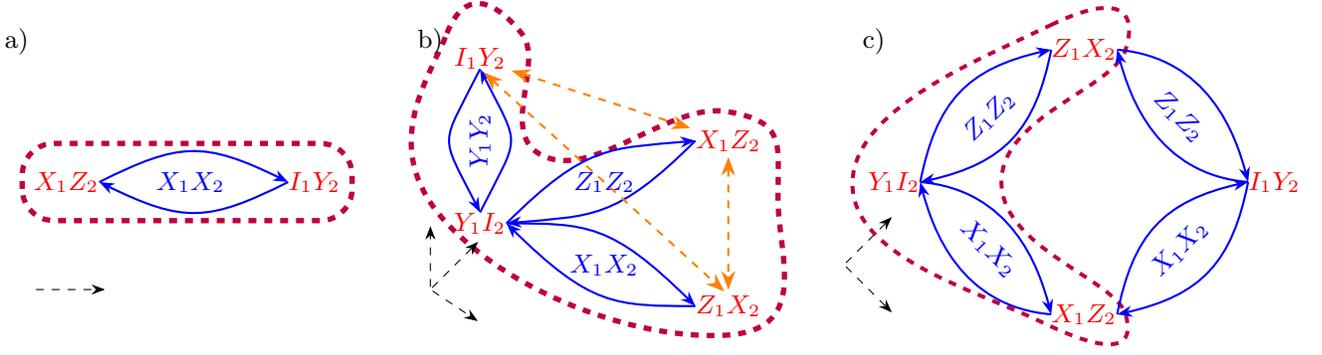

\subsection{Partial Encoding of Single-Qubit channels} \label{subsec: 1-qubit_channel_section}

In the presence of single-qubit noise, the Pauli strings from Eqs. \eqref{system_pauli_channel} and \eqref{qc_channels} simplify to Pauli operators, ${P} \in \{{X}, {Y}, {Z}, {I}\}$. When a system channel, ${P}_i$, is implemented as a gate on a quantum circuit, the intrinsic hardware noise probabilistically composes with this encoded operation and generates an effective weighted sum of Pauli channels within the same single-qubit Pauli group. Depending on the relation between the target system channels and the intrinsic noise channels, this process can lead to two qualitatively distinct cases:
\begin{enumerate}
    \item \emph{Complete encoding of a system channel:}
        This occurs when the intrinsic noise converts the encoded gate into an effective channel that is not present in the system. In such a scenario, we completely encode the channels present in the system and partially correct the resulting channels that arise from the effects of intrinsic noise.
    \item \emph{Partial encoding of a system channel: }
        This occurs when the intrinsic noise converts the encoded gate into effective channels that are partly or completely present in the set of system channels. In such a scenario, we employ an iterative partial encoding scheme to systematically encode all the system channels onto the quantum computer.
    %\item \emph{Partial error correction:} This is a carefully selected process that eliminates any unwanted channels that are encoded on the quantum circuit. {\color{black} To do this, one can program the quantum error correction cycle and obtain the right family of gadgets using convex optimization to mimic the dissipation channels of a target system~\cite{dambal2026programmable}.}
\end{enumerate}

The effective channel implemented in the quantum circuit can also be expressed as a Pauli channel. We define the residual channel, or residue, as the difference between the target system-channel weights and the weights generated by the effective encoded channel. The residue is then fed back into the target weights, and the procedure is repeated until the residual mismatch falls below a chosen tolerance or until the remaining mismatch must be handled by mitigation or partial error correction. {%\color{violet}
For instance, if the open system displays dissipation and dephasing in its dynamics, and the partial encoding algorithm over encodes dephasing at the end of its iterations, then the recovery process of quantum error correction must only correct for this channel. This allows desired noise channels to be encoded as part of the simulation and instructs error correction protocols to correct either the remaining unwanted noise or ones that do not mimic a target open system. In Section \ref{sec:results}, we demonstrate the emergence of such a scenario as part of our algorithm and motivate this requirement. Further details on this partial error correction procedure can be found in Ref.~\cite{dambal2026programmable}.
}
For brevity, and because two-qubit gate noise is often the dominant calibrated error source in gate-based processors, we elaborate on the procedure for two-qubit channels in this manuscript.

\subsection{Partial Encoding of Two-Qubit Channels}\label{subsec:two_qubit_encoding}
In many gate-based quantum processors, two-qubit gates are among the dominant contributors to circuit error.
Accordingly, one often characterizes circuit depth and benchmarks quantum algorithms by counting the number of CNOT or other entangling gates that are implemented.
These implementations are usually imperfect and noisy, and we can write the gate implementation channel as in Eq.~\eqref{qc_channels} with $n=2$ qubits. Under two-qubit noise, the Pauli group contains 16 channels. Analogous to the arguments presented in Sec.~\ref {subsec: 1-qubit_channel_section}, the encoding of a two-qubit system channel as a unitary gate triggers the two-qubit intrinsic noise channels. This results in an effective mapping to a new channel that may or may not be present in the system. Concretely, this mapping is given by
\begin{equation}
    \{w_i,{P}_i\}_S \xrightarrow{\{w_j,{P}_j\}_I} \{\tilde{w},\tilde{P}\}_{\text{eff}},
\end{equation}
where the sets $\{w_i, {P}_i\}_S$ and $\{w_j, {P}_j\}_I$ are the weighted Paulis from Eqs.~\eqref{system_pauli_channel} and~\eqref{qc_channels}. These mappings are determined by products of Pauli matrices. For single-qubit Pauli operators, $\sigma_a\sigma_b=\delta_{ab}I+i\sum_c\epsilon_{abc}\sigma_c$, and the commutator satisfies $[\sigma_a,\sigma_b]=2i\sum_c\epsilon_{abc}\sigma_c$. Thus products of Pauli strings remain Pauli strings up to an irrelevant global phase. Consequently, repeated multiplication by any intrinsic Pauli-noise set $\{{P_j}\}_I \subseteq \{{X}, {Y}, {Z}, {I}\}^{\otimes 2}$ generates a closed orbit inside the two-qubit Pauli group. Physically, if we encode a system channel from $\{{P_i}\}_S \subseteq \{{X}, {Y}, {Z}, {I}\}^{\otimes 2}$, the intrinsic noise induces a closed channel cluster, analogous to a decoherence-free subspace (DFS) in channel space, because all repeated Pauli products remain inside the generated orbit.

Fig.~\ref{fig:braiding-combined}a) shows an example of the two-qubit bit-flip noise ${X}_1{X}_2$ generating a closed map between the ${X}_1{Z}_2$ and ${I}_1{Y}_2$ system channels. Encoding either of the channels under the bit-flip noise will confine the iterative process within this subspace and will not produce new channels. This partial encoding scheme may reduce the amount of error correction required in later stages of the algorithm. Fig.~\ref{fig:braiding-combined}b) gives an example of the intrinsic noise $\{{X}_1{X}_2, {Y}_1{Y}_2, {Z}_1{Z}_2\}_I$ generating a closed map between the elements of the set $\{{Y}_1{I}_2, {Z}_1{X}_2, {X}_1{Z}_2, {I}_1{Y}_2\}_S$ of system channels. In each of these examples, one can identify a node that branches out to multiple channels. This node is the channel encoded as a gate in the quantum circuit and maps to itself under the ${I}_1{I}_2$ intrinsic noise with probability equal to the corresponding magnitude. An example of such a node is ${Y}_1{I}_2$ in Fig.~\ref{fig:braiding-combined}b) where we see three channels branching out of it as indicated by the blue arrows. Note that these branches are reversible up to a global phase. This phase can be ignored since they cancel out in Eq.~\eqref{system_pauli_channel}. The yellow arrows are branches that correspond to alternate node selections and reinforce the generation of a closed channel cluster with all-to-all mappings of the system channels.

Having defined a node and recognized its importance within an iteration of the algorithm, we define a $d_b$-dimensional braiding shown by the purple dashed lines in each subfigure of Fig.~\ref{fig:braiding-combined}. This braiding process can be viewed as a thread that originates from one node and binds all channels that are immediately connected to it by one application of the intrinsic noise. The dimension $d_b$ is the number of channels covered by this one-step process, excluding the node. We also define $d_c$ as the number of distinct Pauli channels in the full closed channel cluster generated by repeated application of the intrinsic noise. The one-step braiding dimension $d_b$ need not equal the full closed-cluster dimension $d_c$.

As a simple example, consider the generation of a DFS under the noise $\{{X}_1{X}_2, {Z}_1{Z}_2\}$, in Fig.~\ref{fig:braiding-combined}c). A node in this figure maps to two other channels as shown by the $d_b=2$ dimensional braiding. However, the intrinsic noise channels create a cyclic map between $4$ system channels, all of which are not present in a single realization of a braiding. We define the dimension, $d_c$, of this DFS as the number of elements present in it and note that $d_c=4 \neq d_b=2$. In particular, as we generalize to multiple intrinsic noise channels, we find that $d_c \geq d_b + 1$ where the equality holds for intrinsic noise channel sets that contain symmetric elements and generate all-to-all mappings in the DFS. Examples of such symmetric sets are the limiting sets, $\{{X}_1{X}_2\} / \{{Y}_1{Y}_2\} / \{{Z}_1{Z}_2\}$ and $ \{{X},{Y}, {Z}, {I}\}^{\otimes 2}$ where the identity channels are implied. Fig.~\ref{fig:braiding-combined}b) demonstrates one such instance of an $d_c=d_b + 1$ DFS structure generated by $3$ symmetric set of intrinsic noise channels. Consequently, we can categorize the DFSs generated into two classes: (i) $d_c=d_b+1$ and (ii) $d_c>d_b + 1$.  

{\color{black} In (i) when the intrinsic noise channels of a quantum processor are well-structured, then they generate a DFS among the available system channels. We define "well-structured" as the intrinsic noise channels that generate an all-to-all mapping (c.f. Figs. \ref{fig:braiding-combined}a and \ref{fig:braiding-combined}b) within the system channel subspace and do not trigger the formation of a channel that is not present within this subspace, i.e. a stray channel. In such a scenario, every step of our algorithm systematically encodes all dissipation channels with various weights without leaking into a stray channel and, as a result, generates a DFS with dimension $d_c$. Since the all-to-all mapping generates the braiding set that equals the DFS (upto an additive constant of 1 by defintion), $d_c = d_b + 1$. 

However, in (ii) when a given set of intrinsic noise channels does not generate all-to-all mapping, only some of the system channels are encoded at each step. As a result, the adaptive encoding algorithm that we present in Sec. \ref{subsec:adaptive_encoding} selects an appropriate new node at every step for ensuring convergence. Since the Pauli group is closed under multiplication, the new nodes facilitate the eventual formation of a larger DFS. In such a scenario, $d_c > d_b + 1$ as seen in Fig. \ref{fig:braiding-combined}c. If $d_c \gg d_b$, then this implies that there exist multiple braiding sets within a DFS. Since an adaptive encoding algorithm identifies an appropriate node at every step and encodes channels within a braiding set, this increases the number of steps required for the algorithm to converge. 

To quantify this effect without invoking a thermodynamic ensemble, we introduce the heuristic braiding-complexity measure $\mathcal{S}_b=\ln\left(\frac{d_c}{d_b+1}\right)$.
This quantity counts, on a logarithmic scale, how many one-step braiding structures are needed to cover a closed channel cluster. For well-structured intrinsic noise with $d_c=d_b+1$, $\mathcal{S}_b=0$. Larger $\mathcal{S}_b$ indicates that adaptive node selection may require more iterations to reduce the residual mismatch.

Finally, if the intrinsic noise generates a Pauli channel outside the target system-channel subspace, that stray channel contributes to the residual mismatch. Such components must be suppressed or mitigated if they are not part of the desired open-system dynamics. One possible route is partial error correction, in which the error-correction cycle is programmed to suppress a fixed set of stray channels while preserving useful dissipative components. Families of operations may include logical resets, logical-frame updates, local gadgets, and feedback primitives. A systematic version of this idea is developed in our subsequent work~\cite{dambal2026programmable}.

The properties and discussion listed so far can be summarized in the two-qubit partial encoding algorithm given in Algorithm~\ref{algo1}.

\begin{algorithm}
\caption{Partial encoding of two- and n-qubit channels}
\label{algo1}
\begin{algorithmic}[1]
\State \textbf{Require: }System channel $\mathcal{E}_{\text{sys}}(\rho) = \sum_k p_k P_k \rho P_k^\dagger$, and intrinsic noise channel $\mathcal{E}_I(\rho) = \sum_i w_i P_i \rho P_i^\dagger$, with $P_k, P_i \in \{X, Y, Z, I\}^{\otimes 2}$
\State  \textbf{Ensure: }$\sum_k p_k = \sum_i w_i = 1$
\State Define $S = \{(p_k, {P}_k) \mid k \in [0,15]\}$ from $\mathcal{E}_{\text{sys}}$
\State Define $I = \{(w_i, {P}_i) \mid i \in [0,15]\}$ from $\mathcal{E}_I$
\State Choose subset $s = \{(p_l, {P}_l)\} \subset S$ \Comment{\textit{This is the node.} One can trivially take $l = 0$.}
%\State Choose subset $s = \{(p_l, {P}_l)\} \subset S$ \Comment{\textit{This is the nodal channel.} Any channel with nonzero residual weight may be selected.}
\For{each iteration $i$}
    \State Encode $s$ on the quantum circuit using Pauli gates
    %\State Compute effective channel resulting from the probabilistic composition of the Pauli gates with the intrinsic noise channels.
    \State Compute the effective channel: $\mathcal{E}_{\text{eff}}(\rho) = \sum_i \tilde{w}_i \widetilde{P}_i \rho \widetilde{P}_i^\dagger$ such that $\tilde{w}_i = p_l w_i$ and $\widetilde{P}_i = {P}_i{P}_l$.
    \State Set $\widetilde{s} = \{(\tilde{w}_k, \widetilde{P}_k) \mid \tilde{w}_k \neq 0\}$
    \State Compute residue: $\Delta^{(i)} = \{(p_k - \tilde{w}_k, {P}_k)\}$
    \State Update $S \gets \Delta^{(i)}$
\EndFor
\State \Return Circuit that encodes $\mathcal{E}_{\text{eff}}(\rho) \approx \mathcal{E}_{\text{sys}}(\rho)$
\end{algorithmic}
\end{algorithm}

\subsection{Generalization to Multiple Qubits}\label{subsec:multi_qubit_encoding}

Algorithm~\ref{algo1}, Fig.~\ref{fig:braiding-combined}, and the ideas developed in Sec.~\ref{subsec:two_qubit_encoding} constitute the foundation of our partial encoding scheme. However, quantum processors consist of multiple qubits and more complex intrinsic noise channels than previously considered. These intrinsic noise channels can vary in strength across qubits and can also be architecture-dependent. More specifically, the ability of a quantum hardware to implement a multi-qubit gate is dependent on its coupling map. Consequently, this affects the kinds of intrinsic noise channels that arise in a quantum processor. To accommodate these complexities, {\color{black} we assume that a given quantum processor implements gates that have support on at most two qubits.} {\color{black} While our partial encoding algorithms remain robust to noisy gates with larger supports, this assumption allows us to probe their consequences more coherently. Moreover, since gates acting on more than two qubits are harder to implement and any multi-qubit gate is eventually decomposed into single- and two-qubit gates~\cite{li2020exponential}, our assumption encompasses a broad family of quantum circuit implementations.}  

With this motivation, we assume that a quantum processor implements only single- and two-qubit gates or transpiles multi-qubit gates into single- and two-qubit gates. As a result, we consider multi-qubit intrinsic noise channels with non-identity noise acting on at most two nearest-neighbor qubits. {\color{black} More specifically, we modify the intrinsic noise channel from Eq.~\eqref{qc_channels} by redefining the support of the Pauli strings onto two nearest-neighbor qubits as,
%The intrinsic noise channel from Eq.~\eqref{qc_channels} can then be written as,
\begin{equation}
    \label{general_two_qubit_gate}
    \mathcal{E}^{(2)}_I(\rho) = \sum_k w_k {P}_k \rho {P}_k^\dagger
\end{equation}
with
\begin{equation}
    \label{Nearest_neighbour_gates}
    {P}_k \in \{{P}\}^N_{nn} \coloneqq \mathbf{P}_{nn}\left[\{X,Y,Z,I\}^{\otimes 2} \otimes {I}^{\otimes (N-2)}\right],
\end{equation}
where $N$ is the number of qubits in the hardware and $\mathbf{P}_{nn}$ permutes the constituent terms in the square brackets such that the first set of tensor product of two operators permutes as a whole with the individual components of the second set of $N-2$ identity operators.} Physically, this means that every two-qubit gate that triggers the intrinsic noise is implemented as a nearest-neighbor instruction. This form is well-motivated because most quantum processors, with various modalities, are able to implement nearest-neighbor two-qubit gates with more efficiency than long-range two-qubit gates. Prominent examples include superconducting quantum processors developed by IBM~\cite{chatterjee2023pairing,zhang2022method}.
%\sd{cite and include more examples}. 
To implement the latter, one usually uses a series of SWAP gates to bring the concerned long-range qubits closer together. This is not only computationally expensive but also harder to implement. For these reasons, we focus our approach on nearest-neighbor gate implementations. 

\begin{figure}[!htb]
   \input{figures/fig_multiple_qubits}
   \caption{a) Extension of the 1D braiding to multiple qubits. This figure shows the formation of mutually exclusive subclusters under the same intrinsic noise $XX$ for two examples of encoded circuits.
   b) Extension of the 3D braiding showing cluster formation by the system channels under the presence of three intrinsic noise channels, $\left\{{X}_1{X}_2, {Y}_1{Y}_2, {Z}_1{Z}_2\right\}$. This assumes identical nearest-neighbor gate implementations.}
   \label{fig:multi_qubit_braiding}
\end{figure}
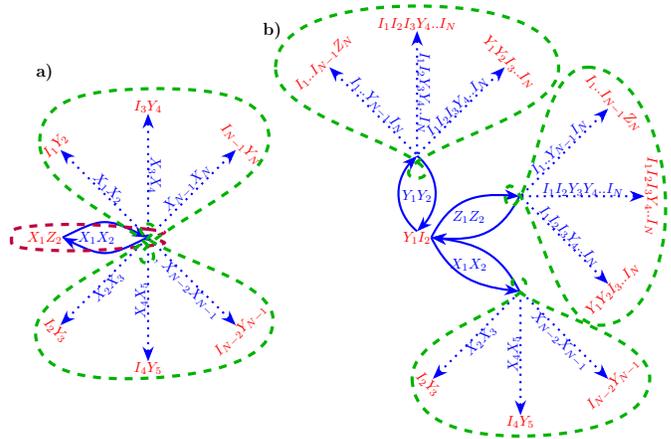

Under the formalism described above, we identify the first term in the square brackets of Eq.~\eqref{Nearest_neighbour_gates} as the constituents of the intrinsic noise used in Algorithm~\ref{algo1}. Since this term consists of only two-qubit Pauli channels, it maps at most two-qubit Paulis at a time from an $N-$qubit Pauli string of the system. The rest of the $N-2$ identity operators act on the remaining Pauli operators of a system channel and leave it unchanged. This means that for an $N$-qubit quantum processor (with $N\%2=0$), a node encodes $(m+1)^{N/2} - 1$ system channels in one iteration, as can be seen in Fig.~\ref{fig:multi_qubit_braiding}. Here, $m$ is the number of intrinsic noise channels that are triggered when a two-qubit gate is implemented. 
As shown in Fig.~\ref{fig:multi_qubit_figures}a), each $ZZ$ gate maps to $YY$ under noise and $ZZ$ otherwise with probabilities dictated by their corresponding weights in the Pauli noise model. The terms contributing to the resultant density matrix come from the tensor products of each of these $YY, ZZ$ channels. As a result, Figs.~\ref{fig:braiding-combined}b) can be extended to Fig.~\ref{fig:multi_qubit_braiding}b) to show these additional encoded channels as subbranches. We see that, under the same intrinsic noise channels (like $XX$ in Fig.~\ref{fig:multi_qubit_braiding}), the system channels form clusters indicated by the green clouds. These clusters are analogous to the DFS presented earlier and are extended versions for multiple-qubit channels. Naturally, for $m=1, N=2$, this reduces to Fig.~\ref{fig:braiding-combined}b). 

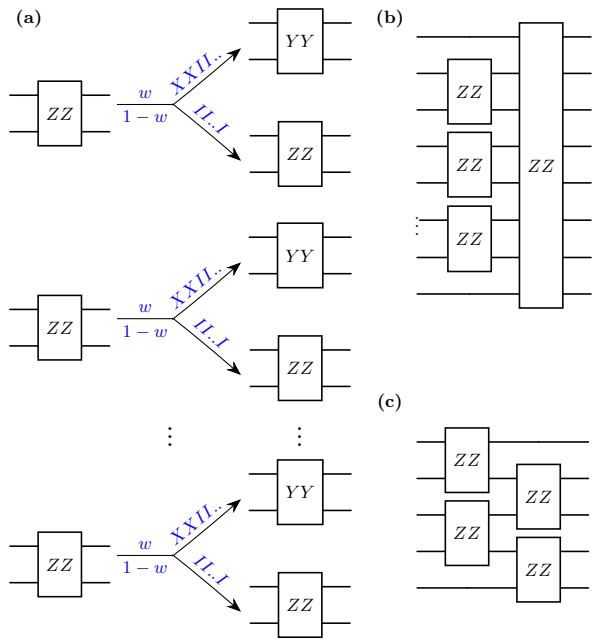
\begin{figure}[!htb]
   \input{figures/fig_quantum_circuits}
   \vspace{-10pt}
   \caption{Examples of two-qubit partially encoded gates:
   (a) (Nearest-neighbor gates) An example of how the encoded node $ZZ..Z$ is transformed under the two-qubit intrinsic noise $XX$.
    (b–c) Two examples of circuits where the nearest-neighbor gate implementation is relaxed. This gives rise to encoded Pauli strings that are different from those obtained in (a) under the same intrinsic noise $XX$.}
   \label{fig:multi_qubit_figures}
\end{figure}

We emphasize that the simplifying constraint imposed in Eq.~\ref{Nearest_neighbour_gates} is not a necessary condition for the implementation of our technique. This is because we can trivially extend it to the more general form in Eq.~\ref{general_two_qubit_gate} by allowing the noise to act non-locally and producing more system channels in each of the clusters in Fig.~\ref{fig:multi_qubit_braiding}b). With these properties in mind, the Algorithm~\ref{algo1} can be trivially generalized to n-qubits by identically applying the identity operator on other sites as described earlier. This leads to the formation of subclusters analogous to the formation of DFS in two-qubits.

\subsection{Adaptive Encoding Scheme}\label{subsec:adaptive_encoding}

The iterative Algorithm~\ref{algo1} can suffer from overencoding.
This occurs when one or more channels are encoded repeatedly in a quest to converge all residues to $0$. As a result, one or more of them diverge to highly negative values and indicate overencoding of the respective system channels. However, this scenario can be mitigated by adaptively encoding the channels at each step of the iteration. This demands an appropriate selection of nodes that mitigate these divergences. Such a nodal selection alters the order in which we encode the system channels in the circuit. However, it can be shown numerically that the output state remains invariant under this ordering, and we can freely choose a node without worrying about it affecting the final density matrix. 
The adaptive encoding scheme that we employ to mitigate divergences in the $d_c=d_b+1$ braiding structure is shown in Algorithm~\ref{adaptive_encoding}. 

\begin{algorithm}
    \caption{Adaptive Encoding Scheme $(d_c=d_b+1)$}
    \label{adaptive_encoding}
\begin{algorithmic}[1]
\State \textbf{Require: }System channel $\mathcal{E}_{\text{sys}}(\rho) = \sum_k p_k P_k \rho P_k^\dagger$, and intrinsic noise channel $\mathcal{E}_I(\rho) = \sum_i w_i P_i \rho P_i^\dagger$, with $P_k, P_i \in \{X, Y, Z, I\}^{\otimes 2}$
\State  \textbf{Ensure: }$\sum_k p_k = \sum_i w_i = 1$
\State Define $S = \{(p_k, {P}_k) \mid k \in [0,15]\}$ from $\mathcal{E}_{\text{sys}}$
\State Define $I = \{(w_i, {P}_i) \mid i \in [0,15]\}$ from $\mathcal{E}_I$
\State Define a tolerance $\text{tol}=0.1$ that sets the residual-mismatch threshold.
\While{$\max_k |p_k| > \text{tol}$ and $\max_k p_k>0$}
    \State \textbf{Adaptive node selection:} Pair the largest positive residual system Pauli channel with the largest intrinsic-noise Pauli component that can generate it. Choose as the node the Pauli string that produces this system channel under that intrinsic-noise component.
    \State Define the node subset $s = \{(p_l, {P}_l)\}$ as above.
    \State Encode $s$ on the quantum circuit using Pauli gates.
    \State Compute the effective channel generated by probabilistically composing the encoded Pauli gates with the intrinsic noise channel:
    \Statex \hspace{1.5em}$\mathcal{E}_{\text{eff}}(\rho)=\sum_i \tilde{w}_i\widetilde{P}_i\rho\widetilde{P}_i^\dagger$, with $\tilde{w}_i=p_lw_i$ and $\widetilde{P}_i=P_iP_l$ up to a global phase.
    \State Set $\widetilde{s}=\{(\tilde{w}_k,\widetilde{P}_k)\mid \tilde{w}_k\neq0\}$, assigning $\tilde{w}_k=0$ to Pauli strings absent from $\widetilde{s}$.
    \State Compute residue: $\Delta^{(i)}=\{(p_k-\tilde{w}_k,{P}_k)\}$.
    \State Update $S\gets\Delta^{(i)}$.
\EndWhile
\State \Return Circuit that encodes $\mathcal{E}_{\text{eff}}(\rho)\approx\mathcal{E}_{\text{sys}}(\rho)$.
\end{algorithmic}
\end{algorithm}

\subsection{Bounds on residue distance}
\label{sec:bound}

Here, we formalize the discussion and algorithms introduced above and derive a practical bound on the residues between the system decoherence channels and those implemented on a quantum processor. Our approach employs the state–channel duality based on the Choi–Jamio{\l}kowski isomorphism~\cite{choi1975completely,jamiolkowski1972linear,jiang2013channel}.

Consider a composite Hilbert space $\mathcal{H}_1 \otimes \mathcal{H}_2$ with $d=\text{dim}(\mathcal{H}_1)=\text{dim}(\mathcal{H}_2)$ and a maximally entangled state
\begin{equation}
    |\Omega\rangle = \frac{1}{\sqrt{d}}\sum_{i=1}^d |i\rangle \otimes |i\rangle,
\end{equation}
where $\{\ket{i}\}_{i=1}^{d}$ forms a complete basis in $\mathcal{H}_1$ or $\mathcal{H}_2$. For a CPTP map $\mathcal{E}:\mathcal{H}_1\to\mathcal{H}_1$, the corresponding Choi state is defined as
\begin{equation}
    \label{choi_matrix}
    J(\mathcal{E}) = (\mathcal{E} \otimes \mathcal{I})(|\Omega \rangle \langle \Omega |),
\end{equation}
where $\mathcal{I}$ is the identity operation on $\mathcal{H}_2$. The action of $\mathcal{E}$ on a state $\rho$ is then given by (see Appendix~\ref{app:channel_state_duality} for details)
\begin{equation}
\label{channel_state_duality}
    \frac{1}{d}\mathcal{E}(\rho) = \text{Tr}_2\big[(\mathbb{I} \otimes \rho^{\top})J(\mathcal{E})\big],
\end{equation}
where $\mathbb{I}$ is the identity operator and $\top$ denotes matrix transpose.

The system channel and the effective channel implemented on a quantum processor define two CPTP maps acting on $\mathcal{H}_1$. Our goal is to match these two channels, taking the system channel as the target. We denote these maps by $\mathcal{E}_{\text{sys}}$ and $\mathcal{E}_{\text{eff}}$, respectively. We use the operator distance defined by the Schatten $p$-norm~\cite{WatrousBook18,horn2012matrix}, referred to as the \textit{Schatten $p$-distance} (with $p=2$ reducing to the Hilbert–Schmidt distance),
\begin{align}
    \left\|A-B\right\|_p := \left(\text{Tr}\left[\Big((A-B)^{\dagger}(A-B)\Big)^{p/2}\right]\right)^{1/p}.
\end{align}

We then obtain bounds on the Schatten $p$-distance between the system channel and the effective channel on the quantum processor (see Appendix~\ref{app:bound} for the proof),
\begin{theorem}
\label{theorem}
\normalfont Let $J(\mathcal{E}_{\text{sys}})$ and $J(\mathcal{E}_{\text{eff}})$ be the Choi states for the CPTP maps $\mathcal{E}_{\text{sys}}$ and $\mathcal{E}_{\text{eff}}$, respectively. Then for any quantum state $\rho$ acting on a $d$-dimensional Hilbert space, the Schatten $p$-distance between $\mathcal{E}_{\text{sys}}(\rho)$ and $\mathcal{E}_{\text{eff}}(\rho)$ can be bounded as 
\begin{equation}
    0 \leq \left\|\mathcal{E}_{\text{sys}}(\rho)-\mathcal{E}_{\text{eff}}(\rho)\right\|_p \leq d\, \left\|J(\mathcal{E}_{\text{sys}})-J(\mathcal{E}_{\text{eff}})\right\|_p~(\forall \rho).
\label{final_bound}
\end{equation}
\end{theorem}
Theorem~\ref{theorem} provides an alternative demonstration of the state–channel duality through the Choi–Jamio{\l}kowski isomorphism, as it establishes the one-to-one correspondence between $J(\mathcal{E})$ and $\mathcal{E}$ defined by the isomorphism. However, this theorem offers a more \textit{practical} viewpoint, with its primary advantage stemming from the use of the Schatten $p$-norm, which provides a versatile framework for analyzing properties such as numerical stability and spectral sensitivity. For example, when $p=2$, the SWAP test can be employed to evaluate the state overlap, thereby yielding the Hilbert--Schmidt distance between two quantum states~\cite{cincio2018learning}. In particular, from Eq.~\eqref{final_bound}, if 
$\norm{J(\mathcal{E}_{\text{sys}}) - J(\mathcal{E}_{\text{eff}})}_{p} \in \mathcal{O}(\varepsilon)$ for arbitrarily small $\varepsilon > 0$, then the channels $\mathcal{E}_{\text{sys}}$ and $\mathcal{E}_{\text{eff}}$ differ by at most $\mathcal{O}(\varepsilon)$, i.e., they are asymptotically close. Consequently, the upper bound $\norm{J(\mathcal{E}_{\text{sys}}) - J(\mathcal{E}_{\text{eff}})}_{p}$ can serve as a cost function, providing a practical and effective metric to achieve perfect partial encoding by iteratively evaluating the efficiently computable Schatten $p$-distance between the two Choi states.

\section{Results and Discussion}\label{sec:results}

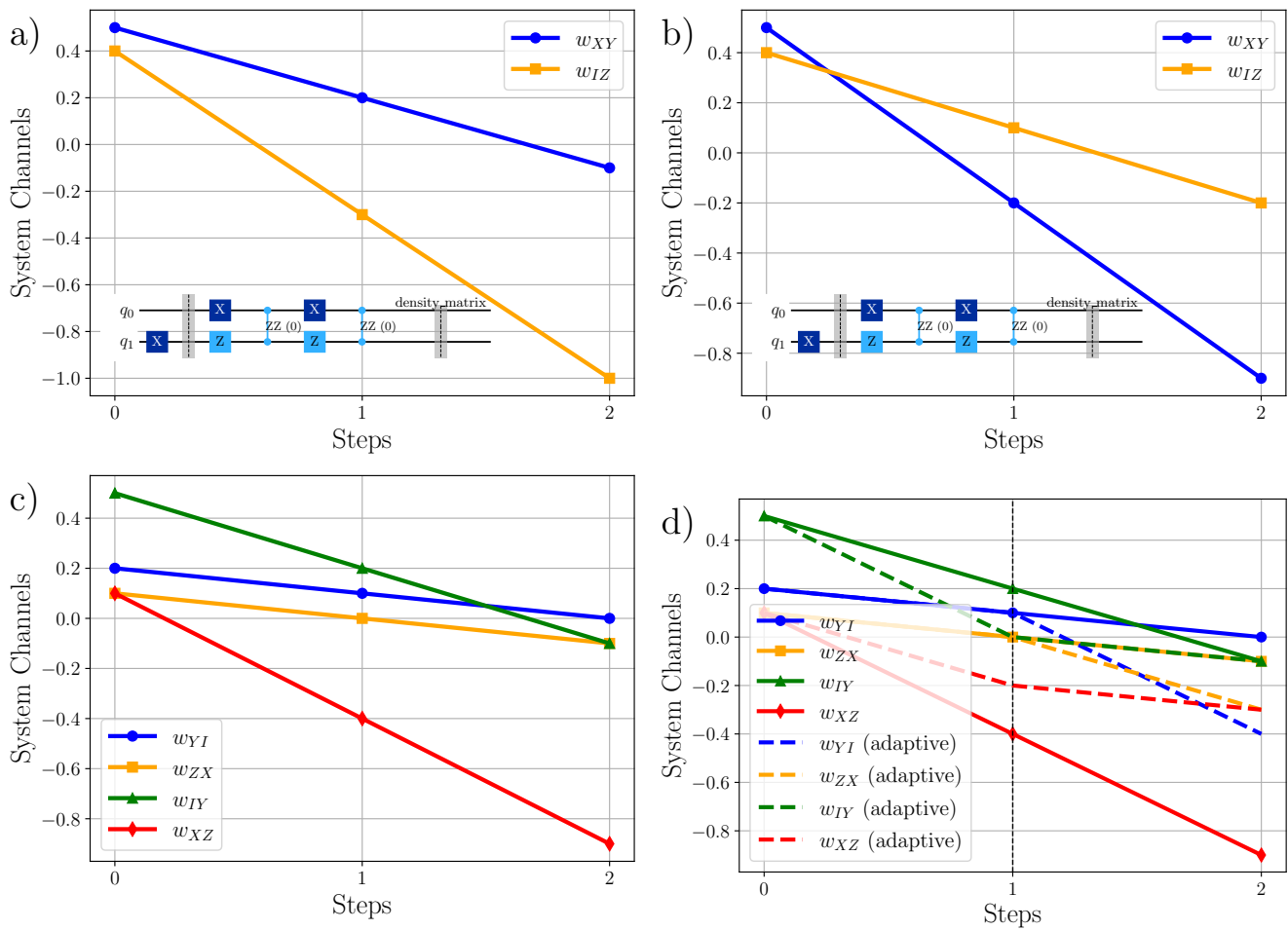
\begin{figure*}[t]
  \begin{tikzpicture}
    \node[inner sep=0] (imgA) {\includegraphics[width=0.49\linewidth]{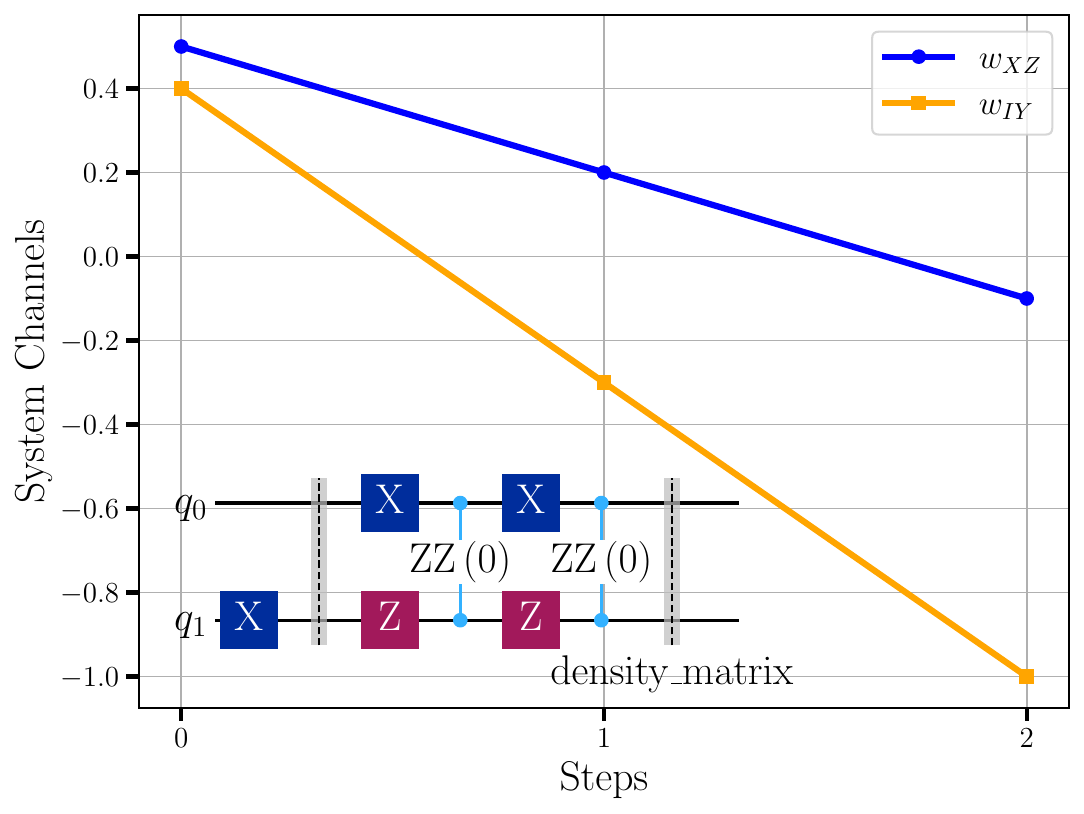}};
    \node[panel label, xshift=2pt, yshift=-5pt] at (imgA.north west) {\small a)};
    %\draw[dashed, thick] ([yshift=0pt]imgA.south west) -- ([yshift=0pt]imgA.south east);
  \end{tikzpicture}
  \begin{tikzpicture}
    \node[inner sep=0] (imgb) {\includegraphics[width=0.49\linewidth]{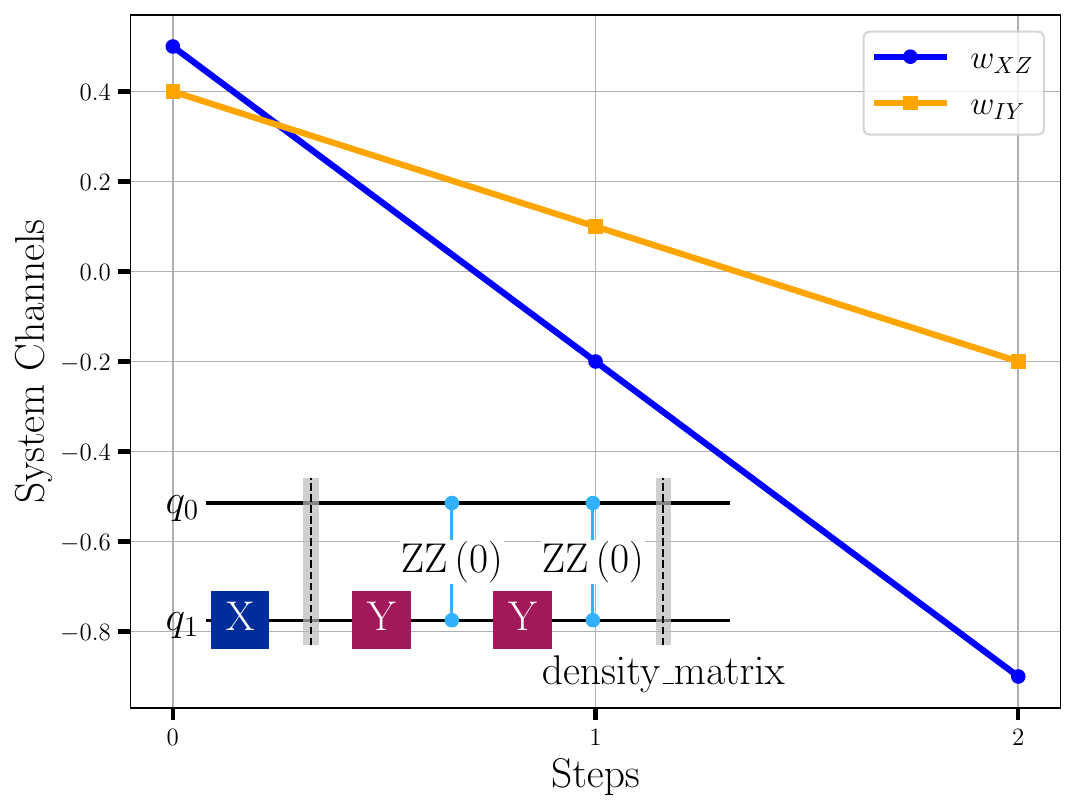}};
    \node[panel label, xshift=2pt, yshift=-5pt] at (imgb.north west) {\small b)};
  \end{tikzpicture}
  \begin{tikzpicture}
    \node[inner sep=0] (imgc) {\includegraphics[width=0.49\linewidth]{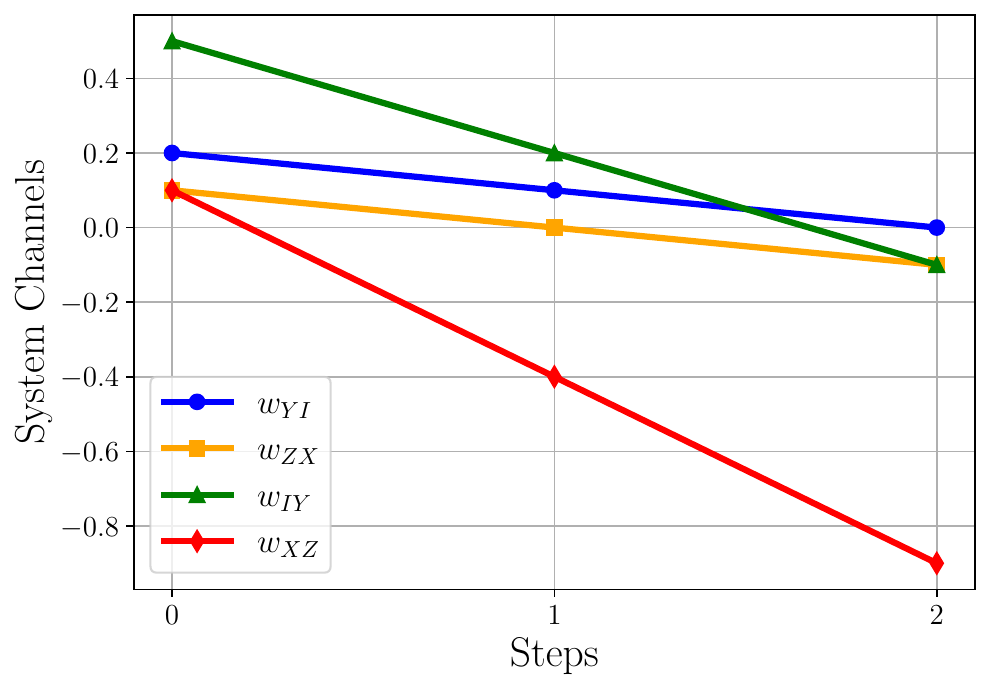}};
    \node[panel label, xshift=2pt, yshift=-5pt] at (imgc.north west) {\small c)};
  \end{tikzpicture}
  \begin{tikzpicture}
    \node[inner sep=0] (imgd) {\includegraphics[width=0.49\linewidth]{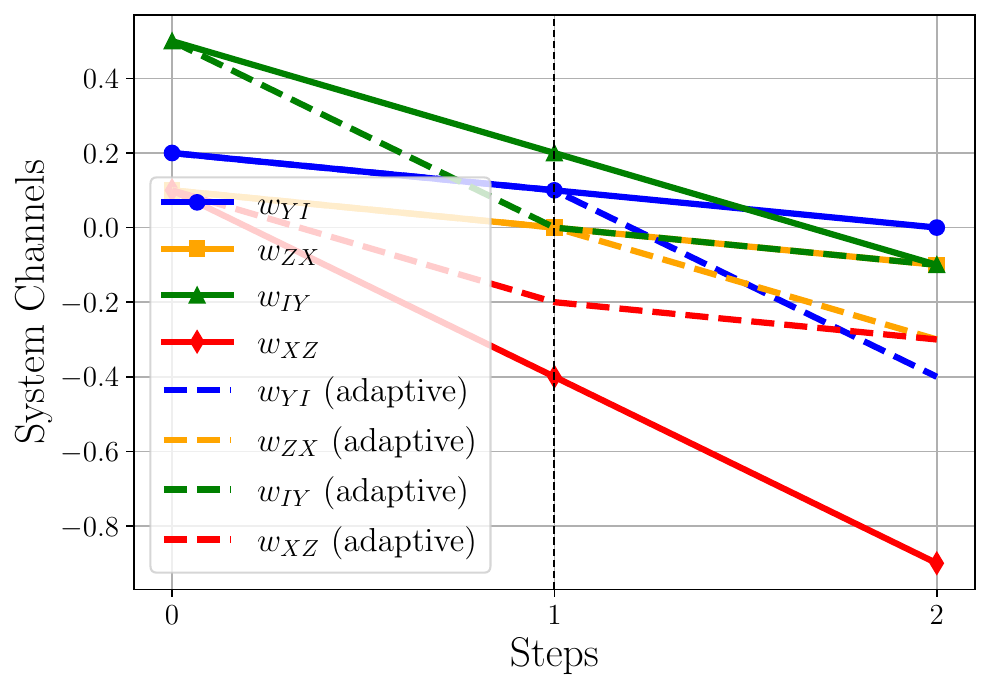}};
    \node[panel label, xshift=2pt, yshift=-5pt] at (imgd.north west) {\small d)};
  \end{tikzpicture}
    \caption{Convergence of the partial encoding algorithms:
    a) Simultaneous convergence of the residuals for the $XZ$ and $IY$ system channels when $XZ$ is explicitly encoded (inset).
    b) Explicitly encoding $IY$ likewise drives simultaneous convergence of both channels.
    In both a) and b), the residuals of the other $16$ non-target two-qubit Pauli channels remain negligible, indicating that under $XX$ intrinsic noise the $XZ$ and $IY$ channels map into each other and define a closed channel cluster.
    c) Similar convergence behavior is observed for the set $\{YI, ZX, IY, XZ\}$ in the presence of intrinsic noise $\{XX, YY, ZZ\}$; over-encoding manifests as residuals diverging to large negative values.
    d) An adaptive scheme that selects which system channel to encode at each step---here choosing $XZ$ and $ZX$ (insets omitted for clarity)---mitigates overencoding and improves the residual mismatch in this example.
    }
    \label{channel_reductions}
\end{figure*}

With the background ideas and algorithms outlined above, we analyze the convergence of our algorithm and benchmark it with well-understood classically simulated systems. In Sec.~\ref{subsec:convergence}, we analyze the convergence of our proposed algorithms, and in Sec.~\ref{subsec:benchmarking}, we benchmark these algorithms with coherent-dissipative energy transport in molecular aggregates to demonstrate their applications in real open quantum systems.

\subsection{Convergence of the partial encoding algorithm}
\label{subsec:convergence}

We begin by demonstrating the convergence of the encoding of a system channel on a quantum computer using its intrinsic noise. Consider the system channels described as follows:
\begin{equation}
    \label{XZ_sys_channel}
    \mathcal{E}_{\text{sys}}(\rho) = w_{XZ} XZ \rho XZ + w_{IY} IY \rho IY
\end{equation}
and the intrinsic noise channel of a quantum computer as,
\begin{equation}
    \label{example_intrinsic_1}
    \mathcal{E}_I(\rho) = w_{XX} XX \rho XX + (1-w_{XX}) \rho.
\end{equation}
Equation~\eqref{example_intrinsic_1} says that whenever a system channel, $XZ$ or $IY$ from Eq.~\eqref{XZ_sys_channel} (or any other channel), is encoded on a quantum circuit as a two-qubit gate, then the resultant density matrix is affected by the intrinsic noise channel $XX$ with probability $w_{XX}$ and stays the same with probability $(1-w_{XX})$. In Fig.~\ref{channel_reductions}a), we begin by encoding the $XZ$ channel and observe simultaneous reduction of both system channels in every iteration. To emphasize that these channels form a closed channel cluster, we instead encode only the $IY$ channel in Fig.~\ref{channel_reductions}b) and observe that the channels converge in a similar manner. The differences in slopes arise from the weights associated with each channel and do not change the logic of the process. This shows that the partial encoding process, under the $XX$ intrinsic noise, does not create a channel outside the subspace of the system's channel space. 

We now extend our analysis by introducing three intrinsic noise channels in the form of $\{{X}{X}, {Y}{Y}, {Z}{Z}\}$ and repeat the partial encoding process for the system channels $\{{Y}{I}, {X}{Z}, {Z}{X}, {I}{Y}\}$. In Fig.~\ref{channel_reductions}c), we see a similar simultaneous encoding of the system channels on the quantum computer per iteration. The algorithm truncates at the point when all the system channels have been completely encoded and the residual system channels (y-axis) converge to zero. However, we also observe that the ${X}{Z}$ channel is being exceedingly encoded as indicated by the large negative value at the end of the iteration. To mitigate this, we employ our adaptive encoding scheme described in Algorithm~\ref{adaptive_encoding}. In Fig.~\ref{channel_reductions}d), we see that the divergences to highly negative values due to this adaptive procedure are successfully tamed. Moreover, we can do better if we define a tolerance below which the iteration can be truncated. This is indicated by the vertical black dotted line and highlights that all channels are encoded within a tolerance of $= 0.1$, allowing the iteration to be truncated at the cost of not completely encoding all the channels. The circuits used in all plots are provided as insets.

\subsection{Benchmarking the algorithm against classical simulations of systems}\label{subsec:benchmarking}

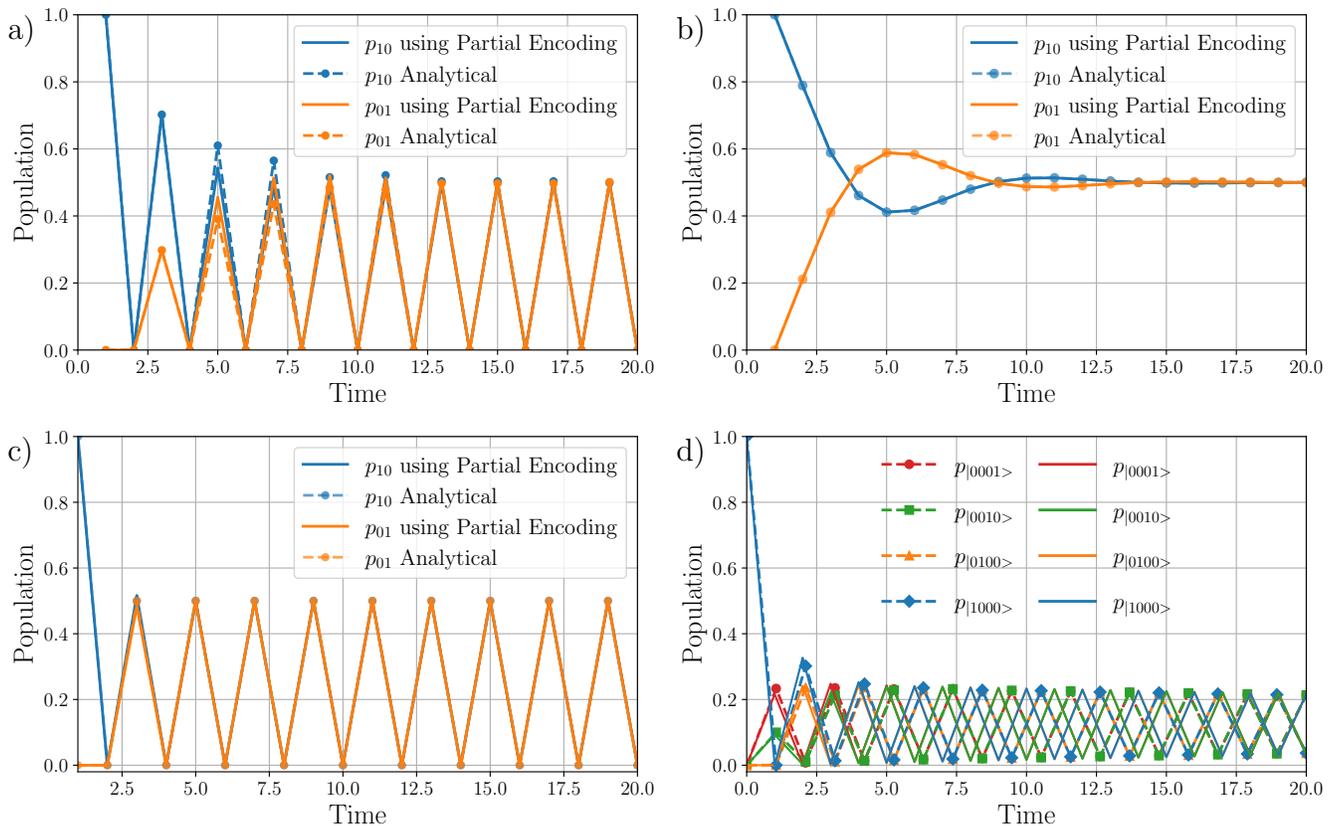
\begin{figure*}[t]
    \centering
    % ---- First row: two subfigures ----
  \begin{tikzpicture}
    \node[inner sep=0] (imga) {\includegraphics[width=0.49\linewidth]{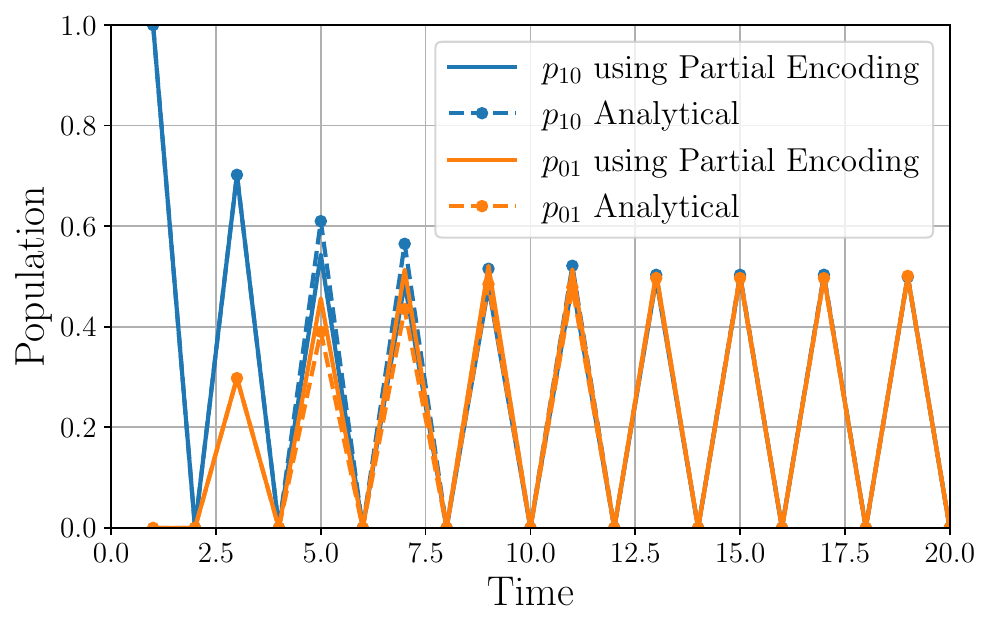}};
    \node[panel label, xshift=0pt, yshift=-5pt] at (imga.north west) {\small a)};
    %\draw[dashed, thick] ([yshift=0pt]imgA.south west) -- ([yshift=0pt]imgA.south east);
  \end{tikzpicture}
  \begin{tikzpicture}
    \node[inner sep=0] (imgb) {\includegraphics[width=0.49\linewidth]{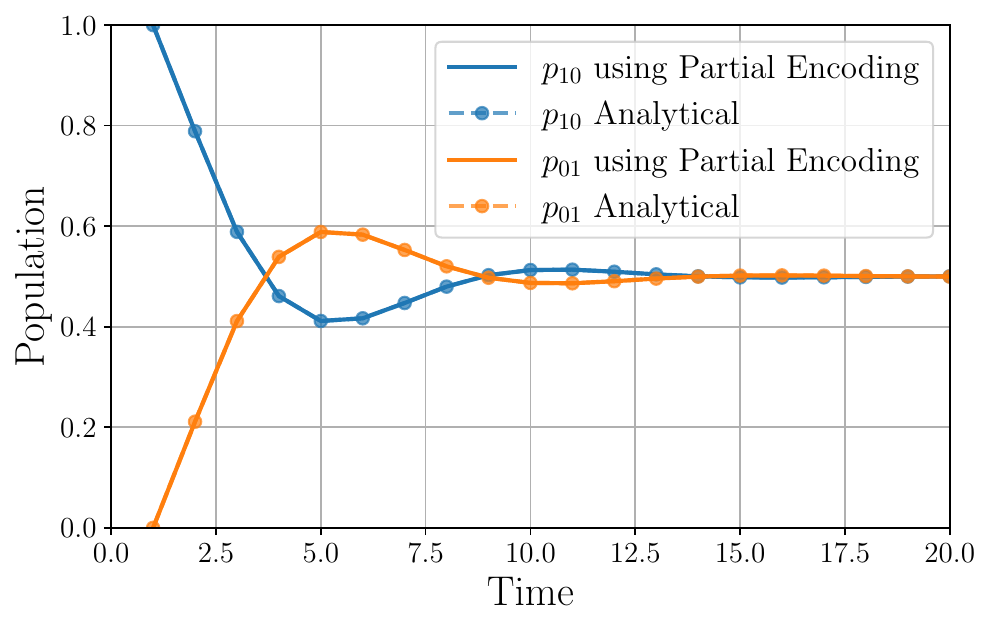}};
    \node[panel label, xshift=0pt, yshift=-5pt] at (imgb.north west) {\small b)};
    %\draw[dashed, thick] ([yshift=0pt]imgA.south west) -- ([yshift=0pt]imgA.south east);
  \end{tikzpicture}
  \begin{tikzpicture}
    \node[inner sep=0] (imgc) {\includegraphics[width=0.49\linewidth]{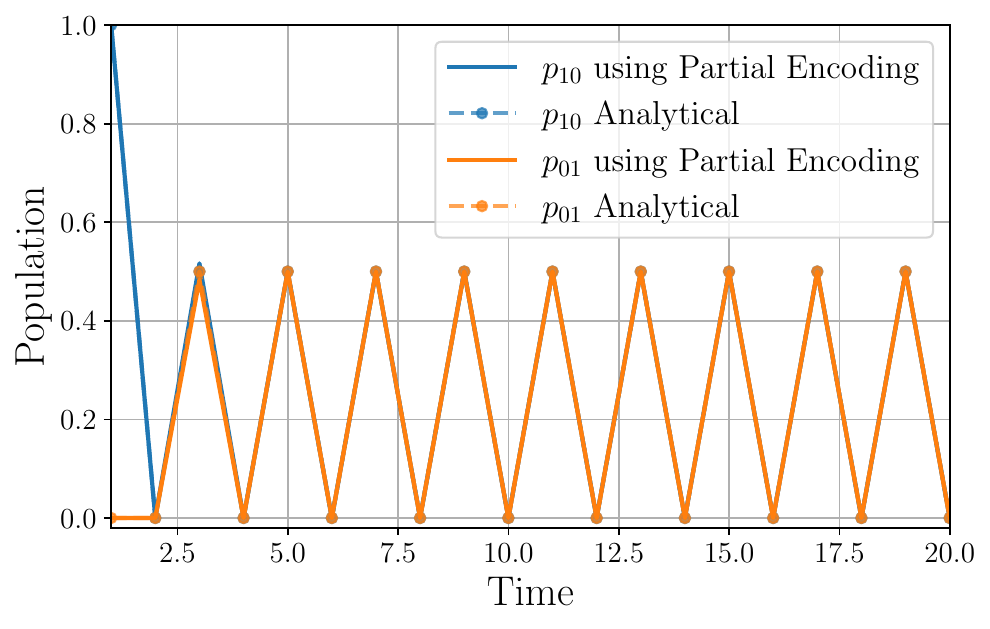}};
    \node[panel label, xshift=0pt, yshift=-5pt] at (imgc.north west) {\small c)};
  \end{tikzpicture}
  \begin{tikzpicture}
    \node[inner sep=0] (imgd) {\includegraphics[width=0.49\linewidth]{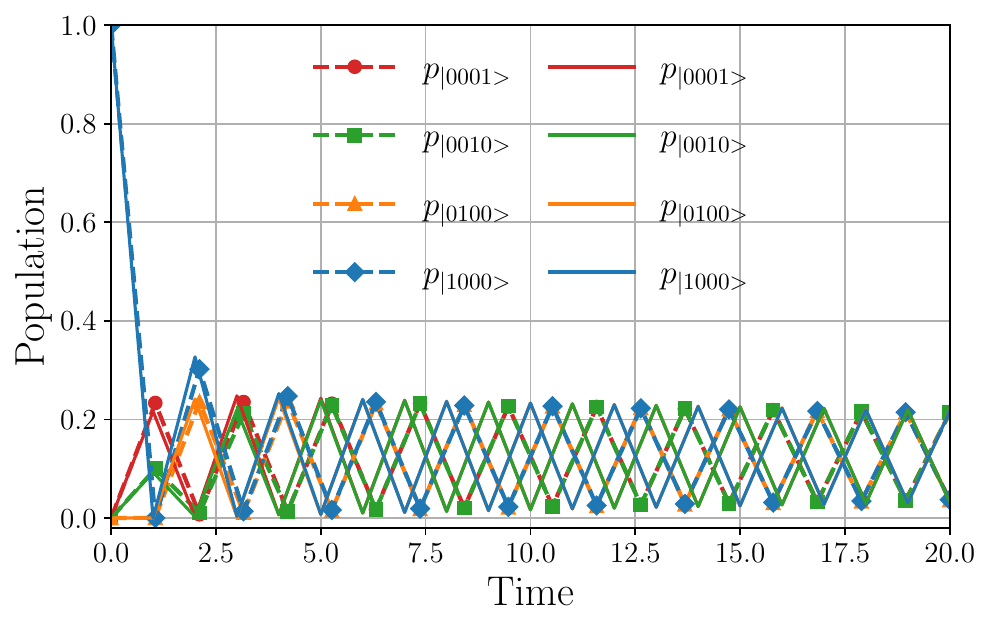}};
    \node[panel label, xshift=0pt, yshift=-5pt] at (imgd.north west) {\small d)};
  \end{tikzpicture} 
    \caption{Benchmarking our algorithm on exciton transport. (a) Two coupled excitons with $\{XZ, IY\}$ decoherence channels. The dotted lines denote analytical calculations, and the solid lines denote the quantum circuit evolution using partial encoding.
    b) shows a similar analysis for $\{ZZ, YY\}$ decoherence channels. In both a) and b), the intrinsic hardware-noise channels are $\{XX, II\}$, and analytical and noise-assisted simulations are in close agreement.
    c) Here we extend the intrinsic noise $\{XX, YY, ZZ\}$ to encode the $\{XZ, IY, ZX, YI\}$ system decoherence channels. d) Extension to four nearest-neighbor coupled excitons with system decoherence channels $\{XZXZ, IYIY, IYXZ, XZIY\}$ under intrinsic noise $\{XX, II\}$. Our noise-assisted simulations closely match analytical predictions; minor deviations stem from Trotterization~\cite{ikeda2023minimum} of the Hamiltonian's nearest-neighbor terms.}
    \label{fig:system_dynamics_benchmark}
\end{figure*}

We benchmark our algorithms with the classically well-understood coherent-dissipative energy transport between molecular aggregates in chemical systems. These molecular aggregates are usually coupled to a phonon bath that introduces incoherent dynamics in their energy transport. We can model this system as a one-dimensional chain of coupled spins with nearest-neighbor coupling. This is given by the Hamiltonian,
\begin{equation}
    \label{Hamiltonian_excitons}
    H = \frac{\omega_0}{2}\sum_{i=1}^N {\sigma}^z_i + g\sum_{i=1}^{N-1} ({\sigma}^+_i {\sigma}^-_{i+1} + {\sigma}^-_i {\sigma}^+_{i+1})
\end{equation}
with open boundary conditions, where $N$ is the number of spins. The complete coherent-dissipative energy transport in this model is usually described by the Lindblad master equation in Eq.~\eqref{Lindblad_equation} with the unitary evolution governed by the Hamiltonian in Eq.~\eqref{Hamiltonian_excitons}. However, using the arguments posed in Sec.~\ref{sec:theory}, we dynamics dynamics of this system more generally be described using the Pauli noise model in Eq.~\eqref{system_pauli_channel}. {\color{black} For example, if we consider a bath that removes coherences from a system in the form of dephasing dynamics, the Lindblad operators describing this process are given by $L_i = \sigma_i^z$. This maps to the the channel $Z_i$ in the Pauli noise model in Eq.~\eqref{system_pauli_channel} In a similar manner, bit flips and bit-phase flips described by $L_i = \sigma^x_i$ and $L_i = \sigma^y_i$ are mapped as $X$ and $Y$ channels. This can then trivially be extended trivially to correlated multi-qubit system channels like $Z_iZ_j$.} \\

In Fig.~\ref{fig:system_dynamics_benchmark}a, we consider a system of two coupled excitons with open-system dynamics described by Eq.~\eqref{XZ_sys_channel} as shown by solid lines. By partially encoding the system channels onto the noisy quantum circuit, we see that the resulting dynamics (dotted lines) match well with the analytically calculated system dynamics. In each time step of the dynamics, we encode the system channels and allow the residues to converge. This process is repeated over all time steps to ensure that all channels are encoded during each step of the evolution. A similar benchmarking is done in Fig.~\ref{fig:system_dynamics_benchmark}b, where the energy transport between the pair of excitons is $\{ZZ, YY\}$ channels. We see that the dotted lines overlap exactly with the solid lines and demonstrate perfect encoding of the system channels using the intrinsic noise $XX$. The algorithm also holds for multiple intrinsic noise channels $\{XX, YY, ZZ\}$. This can be seen in Fig.~\ref{fig:system_dynamics_benchmark}c) where the intrinsic noise simultaneously encodes $\{XZ, IY, ZX, YI\}$ system channels. 

Finally, we extend our system to a chain of multiple excitons with nearest-neighbor coupling. The system channels governing the incoherent energy transport in this model are given as
\begin{equation}
\label{multi_exciton_coherent_dissipative_dynamics}
\begin{split}
\mathcal{E}_{\text{sys}}(\rho) = &w_1 XZXZ \rho XZXZ + w_2 IYIY \rho IYIY\\
&+ w_3 IYXZ \rho IYXZ + w_4 XZIY \rho XZIY.
\end{split}
\end{equation}
In Fig.~\ref{fig:system_dynamics_benchmark}d), we see that in the presence of two-qubit bit-flip noise, the algorithm encodes the system channels well and replicates the coherent-dissipative transport dynamics with minor differences. These differences stem from the Trotterization~\cite{ikeda2023minimum} of the various off-diagonal, many-body coupling terms in the Hamiltonian. 

Finally, we remark that the system and intrinsic noise channels presented in this work serve as examples to demonstrate the algorithms, the formation of clusters, and the braiding structures generated within them. These particular channel choices are not essential to the construction. More generally, larger intrinsic-noise sets can generate larger closed channel clusters containing multiple braiding structures. The existence of these finite channel orbits follows from closure of the Pauli group under multiplication up to phase, including for asymmetric sets of Pauli strings in the system and intrinsic-noise channels.
%These properties make our algorithms robust against all kinds of unital noise affecting a quantum processor, and provide a seamless technique to incorporate noise as a computational resource in open quantum system simulations.
These properties follow from the closure of the Pauli group under multiplication (up to phase) and make the present framework broadly applicable to Pauli-channel descriptions of unital noise affecting a quantum processor, thereby providing a natural route to incorporating intrinsic noise as a computational resource in open quantum system simulation.

\section{Summary}\label{sec:summary}

Quantum computers, from today's NISQ devices to future fault-tolerant architectures, are fundamentally constrained by noise: it can corrupt simulations, degrade output states, and substantially increase the physical-qubit overhead required for error correction. For open-system simulation, there is an additional challenge: standard gate-based quantum architectures implement unitary operations, whereas dissipative open-system dynamics are nonunitary. Rather than suppressing all nonunitary effects to recover perfectly unitary logic, we have asked whether intrinsic noise can be leveraged as a computational resource to emulate such dynamics.
In this work, we answered this question in the affirmative within the restricted Pauli-channel framework by introducing an ancilla-free technique that constructively encodes a system's decoherence channels as Pauli strings and averages the device's inherent noise into the computation, thereby harnessing a resource ubiquitous across physics and chemistry.

%We introduce a technique for simulating open-system dynamics that both harnesses hardware noise and reduces qubit requirements compared to existing methods. Our approach partially encodes system dephasing channels as Pauli strings on a quantum circuit.
Our approach uses partial encoding to activate intrinsic hardware noise and thereby generate an effective CPTP map that approximates the desired dissipative evolution without explicit ancillary dilation. Relative to dilation-based approaches, this can reduce ancillary overhead and may also reduce the amount of error correction required, since only the mismatch between the useful intrinsic dissipation and the target channel needs to be corrected. We further showed that, because the Pauli group is closed under multiplication up to phase, intrinsic noise induces closed clusters of partially encoded channels. These invariant channel sectors are analogous to decoherence-free subspaces (DFSs)~\cite{lidar1998decoherence,lidar1999concatenating,kwiat2000experimental,bacon1999robustness,lidar2001decoherence,wang2013numerical,bacon2000universal,sone2025no} and provide additional structure that can be exploited in noise-assisted open-system simulation.

Throughout this work, we assumed a Pauli noise model describing both the intrinsic hardware noise and the target decoherence channels. Accordingly, the present framework primarily addresses unital CPTP dissipation. It does not directly implement generic non-CPTP dynamics, such as an unregularized Redfield equation, and genuinely memory-kernel non-Markovian dynamics would require additional resources such as memory registers, feedback, or a generalized channel representation. Time-dependent non-Markovian finite-time maps can be treated step by step only when each step is CPTP and admits a sufficiently accurate Pauli-channel representation. Extending the method to non-unital processes, such as amplitude damping ($T_1$ noise), non-Pauli coherent noise, and explicit memory effects 
remains an important direction for future work. More broadly, partial encoding alone does not in general guarantee an arbitrarily small mismatch between the encoded dissipation and an arbitrary target dissipation channel, even within the present Pauli-channel setting. Depending on the interplay between the target and the hardware noise, the protocol may exhibit overencoding or underencoding of specific dissipative components. In this regime, Theorem~\ref{theorem} provides a practical bound on the distance between the target CPTP map and the effective encoded map through the distance between their Choi states, thereby quantifying the residual error without, by itself, guaranteeing that the mismatch can always be made arbitrarily small.

This limitation motivates an important next step that is the subject of follow-up work: a \emph{partial quantum error-correction}~\cite{dambal2026programmable} scheme that selectively suppresses only the unwanted components of the intrinsic noise while preserving the dissipative components useful for simulation. Such an approach would allow one to program a logical CPTP map that more faithfully mimics a desired target dissipation channel and could provide a route toward near-optimal minimization of the residual channel distance characterized by Theorem~\ref{theorem}. By combining partial encoding with selective error correction, we envision a practical route toward more accurate and resource-efficient quantum simulation of open quantum systems on both NISQ and future fault-tolerant devices.

\begin{dataavailability}
The source code and data to generate the figures in the paper are available upon request.
\end{dataavailability}

\begin{acknowledgements}
The research presented in this article was supported by the Laboratory Directed Research and Development (LDRD) program of Los Alamos National Laboratory (LANL). We thank the LANL Institutional Computing (IC) program for access to HPC resources. LANL is operated by Triad National Security, LLC, for the National Nuclear Security Administration of the U.S. Department of Energy (contract no. 89233218CNA000001). A.S. acknowledges the U.S. NSF under Grant No. OSI-2328774, PHY-2425180, and Cooperative Agreement PHY-2019786.
\end{acknowledgements}

\bibliography{qc4oqs.bib}

\appendix

\onecolumngrid

\input{appendix}

\end{document}

%% file: tikzcommand.tex
\usepackage{xcolor}
\usepackage{physics}
\usepackage[outline]{contour} % glow around text

\usepackage{tikz}
\usepackage{circuitikz}
\usepackage{quantikz}
\usepackage{pgfgantt}
\usepackage{siunitx}

\usepackage{pgfplots}
\pgfplotsset{compat=newest}

\usetikzlibrary{3d}
\usetikzlibrary{angles,quotes} % for pic (angle labels
\usetikzlibrary{patterns} % Include this line to enable pattern lib)
\usetikzlibrary{backgrounds}
\usetikzlibrary{snakes}
\usetikzlibrary{calc, decorations.pathreplacing, decorations.markings} %fadings,
\usetikzlibrary{decorations.pathmorphing}

\tikzset{>=latex} % for LaTeX arrowhead

\usetikzlibrary{fit, positioning}
\usetikzlibrary{intersections, patterns, pgfplots.fillbetween}
\usetikzlibrary{arrows, arrows.meta, shapes.geometric}

\usetikzlibrary{hobby} % for ..

\definecolor{tensorblue}{rgb}{0.8,0.8,1}
\definecolor{tensorred}{rgb}{1,0.5,0.5}
\definecolor{tensorpurp}{rgb}{1,0.5,1}

\tikzset{ten/.style={fill=tensorblue}}
\tikzset{tenred/.style={fill=tensorred}}
\tikzset{tengreen/.style={fill=green!50!black!50}}
\tikzset{tenpurp/.style={fill=tensorpurp}}
\tikzset{tengrey/.style={fill=black!20}}
\tikzset{tenorange/.style={fill=orange!30}}
\tikzset{u/.style={fill=blue!20,draw=black}}
\tikzset{w/.style={fill=green!50!black!80,draw=black}}

\pgfdeclarelayer{l1}
\pgfdeclarelayer{l2}
\pgfsetlayers{l1,l2,main}

\definecolor{myred}{HTML}{FF1F5B}
\definecolor{myblue}{HTML}{009ADE}
\definecolor{mygreen}{HTML}{00CD6C}
\definecolor{myorange}{HTML}{F27522}
\colorlet{mucol}{red!90!black}
\colorlet{agg_col}{blue!80!}

\tikzstyle arrowstyle=[scale=1]
\tikzstyle directed=[postaction={decorate,decoration={markings,
    mark=at position .65 with {\arrow[arrowstyle]{stealth}}}}]
\tikzstyle reverse directed=[postaction={decorate,decoration={markings,
    mark=at position .65 with {\arrowreversed[arrowstyle]{stealth};}}}]

\tikzstyle{vector_red}=[->,line width=2.0pt,mucol]
\tikzstyle{vector_blue}=[->,line width=2.0pt,myblue]
\tikzstyle{dashed_red}=[->,line width=2.0pt, mucol, dashed]

\makeatletter
\pgfkeys{%
  /tikz/node on layer/.code={
    \gdef\node@@on@layer{%
      \setbox\tikz@tempbox=\hbox\bgroup\pgfonlayer{#1}\unhbox\tikz@tempbox\endpgfonlayer\egroup}
    \aftergroup\node@on@layer
  },
  /tikz/end node on layer/.code={
    \endpgfonlayer\endgroup\endgroup
  }
}

\def\node@on@layer{\aftergroup\node@@on@layer}
\makeatother

\tikzset{%
  >={Latex[width=2mm,length=2mm]},
            base/.style = {rectangle, rounded corners, draw=black,
                           minimum width=4cm, minimum height=1cm,
                           text centered}, %font=\rmfamil},
  activityStarts/.style = {base, fill=blue!20},
       startstop/.style = {base, fill=red!30},
    activityRuns/.style = {base, fill=green!30, minimum size=2mm},
         process/.style = {draw, fill=orange!15, dashed, inner sep = 5pt, %minimum width=0.5cm, 
                           },%font=\ttfamily},
  lvl1/.style={draw,fill=red!20,rounded corners=0.2cm,inner sep=5pt,node on layer=l1,  minimum size=2mm},
  lvl2/.style={draw,fill=blue!25,rounded corners=0.2cm,inner sep=5pt,node on layer=l2,  minimum size=2mm},
  lvl3/.style={draw=blue,fill=white,dashed,rounded corners=0.25cm,align=flush center,text width=12em,inner sep=4pt,minimum height=1.5cm}, 
  title/.style={node font=\large,  minimum size=1mm},
  myarrow/.style={latex-latex,ultra thick,blue!80},
}

\tikzset{circle with arrow/.style args={#1,#2}{
        circle, draw=none, 
        thick, inner sep=1pt,
        minimum size=12pt, fill=#1,
        font=\sffamily\bfseries\color{#2}
    },
    arrow style/.style={
        ->, line width=1.4pt, % make the arrow thicker
        >=latex,
        white
    }
}

\tikzset{
  panel label/.style={
    anchor=north west,
    font=\footnotesize,
    fill=white,
    inner sep=1.5pt,
    rounded corners=1pt
  }
}

%% file: commands.tex
\newenvironment{dataavailability}
    {\par\begin{center}\textbf{\MakeUppercase{Data Availability}}\end{center}\noindent}
    {\par}

%% file: figures/1D_braiding.tex
\begin{tikzpicture}[baseline, 
every node/.style={inner sep=0pt}, >=Stealth,
scale=1.1, transform shape
]

\pgfdeclarelayer{background}
\pgfsetlayers{background,main}

%---------------------------------------
% panel a
%---------------------------------------

  \node (label1) at (-0.6, 2.2) {a)};
  %--- red Pauli strings
  \node (A) at (0.0, 0.5) {\(\color{red}{  X_{1}  Z_{2}}\)};
  \node (B) at (3.0, 0.5) {\(\color{red}{  I_{1} Y_{2}}\)};

  %--- blue double arrows
  \draw[thick, blue, -{Stealth[length=2mm]}] (A.east) .. controls +(1.,0.5) and +(-1.,0.5) .. (B.west);
  \draw[thick, blue, -{Stealth[length=2mm]}] (B.west) .. controls +(-1.,-0.5) and +(1.,-0.5) .. (A.east);

  %--- blue label in the middle
  \node at ($(A)!0.5!(B)$) {\(\color{blue}{  X_{1}  X_{2}}\)};

  %--- dashed double-headed "braiding" arrow to the left
  % \draw[dashed, -{Stealth[length=1.8mm]}-{Stealth[length=1.8mm]}] (0,-0.5) -- (0.5,-0.5);
  % \draw[dashed, -{Stealth[length=1.8mm]}-{Stealth[length=1.8mm]}, shorten >=1pt, shorten <=1pt]  (0,-0.5) -- (0.5,-0.5);
  \draw[dashed, ->, >={Stealth[length=1.8mm]}, shorten >=1pt, shorten <=1pt] (-0.4,-0.8) -- (0.5,-0.8);

  %--- purple dashed "cloudy" rounded outline
  \node[fit=(A)(B), inner sep=4pt] (box) {}; % invisible fit box
  
  % bumpy rectangular 
  %\draw[thick, purple, dashed, rounded corners=12pt, decorate, decoration={bumps, amplitude=2.3pt, segment length=10pt}]
   % ($(box.south west)+(-0.4,-0.45)$) rectangle    ($(box.north east)+(0.4,0.45)$);

  % smooth ellipse around A and B
  % \node[draw=purple, thick, dashed, ellipse, fit=(A)(B), inner sep=10pt] (oval) {};

% smooth rectangle
 \draw[line width=2, purple, dashed, rounded corners=12pt] ($(box.south west)+(-0.0,-0.2)$) rectangle ($(box.north east)+(0.0,0.2)$);

\iffalse
%“capsule” (rounded rectangle with big corners)
\draw[thick, purple, dashed, rounded corners=20pt]
  ($(box.south west)+(-0.4,-0.45)$)
  -- ($(box.south east)+(0.4,-0.45)$)
  -- ($(box.north east)+(0.4,0.45)$)
  -- ($(box.north west)+(-0.4,0.45)$)
  -- cycle;
\fi

%---------------------------------------
% panel b
%---------------------------------------

  \def\xshift{5}
  \node (label2) at (\xshift-1, 2.2) {b)};

  % dimension label
  \draw[dashed, ->, >={Stealth[length=1.8mm]}, shorten >=1pt, shorten <=1pt] (\xshift-1.0,-0.8) -- (\xshift - 0.4, -1.2);
  \draw[dashed, ->, >={Stealth[length=1.8mm]}, shorten >=1pt, shorten <=1pt] (\xshift-1.0,-0.8) -- (\xshift - 0.4,-0.2);
  \draw[dashed, ->, >={Stealth[length=1.8mm]}, shorten >=1pt, shorten <=1pt] (\xshift-1.0,-0.8) -- (\xshift - 1.0, 0.0);
  
  %--- red Pauli strings
  \node (B1) at (\xshift, 0) {\(\color{red}{  Y_{1} I_{2}}\)};
  \node (B2) at (\xshift+ 3.0, -1) {\(\color{red}{ Z_{1}  X_{2}}\)};
  \node (B3) at (\xshift+ 3.0, 1) {\(\color{red}{ X_{1}  Z_{2}}\)};
  \node (B4) at (\xshift, 2) {\(\color{red}{  I_{1}  Y_{2}}\)};

  \begin{pgfonlayer}{background}
  \draw[line width=2, purple, dashed]
    plot [smooth cycle, tension=0.9]
    coordinates {
      ($(B1)+(-0.4,0)$)
      ($(B4)+(-0.4,0.3)$)
      ($(B4)+(0.5,0.3)$)
      ($(B1)+(0.8, 0.8)$)
      ($(B3)+( 0.0,0.35)$)
      ($(B3)+( 0.6,-0.7)$)
      ($(B2)+( 0.0,-0.35)$)
    };
  \end{pgfonlayer}

  %--- blue label in the middle
  \draw[thick, blue, -{Stealth[length=2mm]}] (B1.east) .. controls +(1., 1.0) and +(-1., -0.1) .. (B3.west);
  \draw[thick, blue, -{Stealth[length=2mm]}] (B3.west) .. controls +(-1.,-1) and +(1., 0.1) .. (B1.east);
  \node at ($(B1)!0.5!(B3)$) {\(\color{blue}{  Z_{1}  Z_{2}}\)};

  %--- blue label in the middle
  \draw[thick, blue, -{Stealth[length=2mm]}] (B1.east) .. controls +(1.,0.0) and +(-1., 1.0) .. (B2.west);
  \draw[thick, blue, -{Stealth[length=2mm]}] (B2.west) .. controls +(-1.,-0.0) and +(1.,-1.0) .. (B1.east);
  \node at ($(B1)!0.5!(B2)$) {\(\color{blue}{  X_{1}  X_{2}}\)};

  %--- blue label in the middle
  \draw[thick, blue, -{Stealth[length=2mm]}] (B1.north) .. controls +(0.5, 1) and +(0.5, -1.0) .. (B4.south);
  \draw[thick, blue, -{Stealth[length=2mm]}] (B4.south) .. controls +(-0.5, -1.0) and +(-0.5, 1.0) .. (B1.north);
  \node[rotate=90] at ($(B1)!0.5!(B4)$) {\(\color{blue}{  Y_{1}  Y_{2}}\)};

  % alternative connection
  \draw[dashed, orange, thick, <->, >={Stealth[length=2.5mm]}, shorten >=2pt, shorten <=2pt] (B4.south) -- (B2.north);
  
  \draw[dashed, orange, thick, <->, >={Stealth[length=2.5mm]}, shorten >=2pt, shorten <=2pt] (B3.south) -- (B2.north);

  \draw[dashed, orange, thick, <->, >={Stealth[length=2.5mm]}, shorten >=2pt, shorten <=2pt] (B4.south east) -- (B3.north west);
  
\iffalse
  %--- purple dashed "cloudy" rounded outline
  \node[fit=(B1)(B2)(B3)(B4), inner sep=4pt] (box) {}; % invisible fit box
  % smooth rectangle
   \draw[thick, purple, dashed, rounded corners=12pt] ($(box.south west)+(-0.0,-0.2)$) rectangle ($(box.north east)+(0.0,0.2)$);
\fi

%---------------------------------------
% panel c
%---------------------------------------

  \def\xshift{10}
  \node (label2) at (\xshift-0.24, 2.2) {c)};

  % dimension label
  \draw[dashed, ->, >={Stealth[length=1.8mm]}, shorten >=1pt, shorten <=1pt] (\xshift-0.6,-0.9) -- (\xshift + 0.0, -1.5);
  \draw[dashed, ->, >={Stealth[length=1.8mm]}, shorten >=1pt, shorten <=1pt] (\xshift-0.6,-0.9) -- (\xshift + 0.0, -0.4);
  
  %--- red Pauli strings
  \node (C1) at (\xshift, 0.5) {\(\color{red}{  Y_{1}  I_{2}}\)};
  \node (C2) at (\xshift+ 2.3, -1.1) {\(\color{red}{  X_{1}  Z_{2}}\)};
  \node (C3) at (\xshift+ 2.3,  2.1) {\(\color{red}{  Z_{1}  X_{2}}\)};
  \node (C4) at (\xshift+ 4.6, 0.5) {\(\color{red}{  I_{1}  Y_{2}}\)};

  \begin{pgfonlayer}{background}
  \draw[line width=1.5, purple, dashed]
    plot [smooth cycle, tension=0.9]
    coordinates {
      ($(C1)+(-0.5,0.0)$)
      %($(C1)+(0.5, 1)$)
      ($(C3)+(-0.4,0.3)$)
      ($(C3)+(0.5,0.0)$)
      ($(C1)+(1.3, 0)$)
      %($(C2)+(-0.3,0.7)$)
      ($(C2)+(0.5,0)$)
      ($(C2)+(-0.5,-0.1)$)
      %($(C1)+(-0.30,-0.4)$)
    };
  \end{pgfonlayer}

  %--- blue label in the middle
  \draw[thick, blue, -{Stealth[length=2mm]}] (C1.east) .. controls +(0.2,   1.0) and +(-1., -0.2) .. (C3.west);
  \draw[thick, blue, -{Stealth[length=2mm]}] (C3.west) .. controls +(-0.2,-1.0) and +(1,0.2) .. (C1.east);
  \node[rotate=45] at ($(C1)!0.5!(C3)$) {\(\color{blue}{  Z_{1}  Z_{2}}\)};

  %--- blue label in the middle
  \draw[thick, blue, -{Stealth[length=2mm]}] (C1.east) .. controls +(1.2, -0.1) and +(-0.1, 0.4) .. (C2.west);
  \draw[thick, blue, -{Stealth[length=2mm]}] (C2.west) .. controls +(-1.2, 0.1) and +(0.1, -0.4) .. (C1.east);
  \node[rotate=-45] at ($(C1)!0.5!(C2)$) {\(\color{blue}{  X_{1}  X_{2}}\)};

  %--- blue label in the middle
  \draw[thick, blue, -{Stealth[length=2mm]}] (C3.east) .. controls +(1.2, -0.1)    and +(-0.1, 0.4) .. (C4.west);
  \draw[thick, blue, -{Stealth[length=2mm]}] (C4.west) .. controls +(-1.2, 0.1) and +(0.1, -0.4) .. (C3.east);
  \node[rotate=-45] at ($(C3)!0.5!(C4)$) {\(\color{blue}{  Z_{1}  Z_{2}}\)};

  %--- blue label in the middle
  \draw[thick, blue, -{Stealth[length=2mm]}] (C2.east) .. controls +(0.2,   1.0) and +(-1., -0.2) .. (C4.west);
  \draw[thick, blue, -{Stealth[length=2mm]}] (C4.west) .. controls +(-0.2,-1.0) and +(1,0.2) .. (C2.east);
  \node[rotate=45] at ($(C2)!0.5!(C4)$) {\(\color{blue}{  X_{1}  X_{2}}\)};

\iffalse
  %--- purple dashed "cloudy" rounded outline
  \node[fit=(B1)(B2)(B3)(B4), inner sep=4pt] (box) {}; % invisible fit box
  % smooth rectangle
   \draw[thick, purple, dashed, rounded corners=12pt] ($(box.south west)+(-0.0,-0.2)$) rectangle ($(box.north east)+(0.0,0.2)$);
\fi

\end{tikzpicture}

%% file: figures/fig_multiple_qubits.tex
\usetikzlibrary{shapes, backgrounds}
\begin{tikzpicture}[baseline, 
scale=0.55, transform shape,
font=\normalsize,
every node/.style={inner sep=0pt}, >=Stealth]

%---------------------------------------
% Panel A
%---------------------------------------

\node (label1) at (0, 4) {\large\bf a)};

% Red Pauli strings
  \node (A) at (0.0, 0) {\color{red} $X_1Z_2$};
  \node (B) at (2.5, 0) {};

  %--- blue double arrows
  \draw[thick, blue, -{Stealth[length=2mm]}] (A.east) .. controls +(1.,0.5) and +(-1.,0.5) .. (B.west);
  \draw[thick, blue, -{Stealth[length=2mm]}] (B.west) .. controls +(-1.,-0.5) and +(1.,-0.5) .. (A.east);

    %--- blue label in the middle
  \node at ($(A)!0.5!(B)$) {\color{blue} $X_1X_2$};

  % Blue dotted arrow\coordinate (Bcenter) at (B);
    \coordinate (Bcenter) at (B);
    \coordinate (tip1) at ($(Bcenter)+(135:3)$);
    \coordinate (tip2) at ($(Bcenter)+(90:3)$);
    \coordinate (tip3) at ($(Bcenter)+(45:3)$);
    
    \draw[thick, blue, dotted, -{Stealth[length=2mm]}] (B) -- ++(135:3) node[midway, above, sloped]{\color{blue}$X_1X_2$} node[pos=1, above, rotate=45]{\color{red}$I_1Y_2$}; 
    \draw[thick, blue, dotted, -{Stealth[length=2mm]}] (B) -- ++(90:3) node[midway, above, sloped]{\color{blue}$X_3X_4$} node[pos=1, rotate=0, anchor=south]{\color{red} $I_3Y_4$};
    \draw[thick, blue, dotted, -{Stealth[length=2mm]}] (B) -- ++(45:3) node[pos=1, rotate=-45, anchor=south] {\color{red}$I_{N-1}Y_N$} node[midway, above, sloped]{\color{blue}$X_{N-1}X_N$};

    \coordinate (tip4) at ($(Bcenter)+(-135:3)$);
    \coordinate (tip5) at ($(Bcenter)+(-90:3)$);
    \coordinate (tip6) at ($(Bcenter)+(-45:3)$);

    \draw[thick, blue, dotted, -{Stealth[length=2mm]}] (B) -- ++(-135:3) node[midway, below, sloped]{\color{blue}$X_2X_3$} node[pos=1, rotate=-45, anchor=north]{\color{red}$I_2 Y_3$};
    \draw[thick, blue, dotted, -{Stealth[length=2mm]}] (B) -- ++(-90:3) node[midway, above, sloped]{\color{blue}$X_4X_5$} node[pos=1, rotate=0, anchor=north]{\color{red}$I_4Y_5$};
    \draw[thick, blue, dotted, -{Stealth[length=2mm]}] (B) -- ++(-45:3) node[midway, below, sloped]{\color{blue}, $X_{N-2}X_{N-1}$} node[pos=1, rotate=45, anchor=north]{\color{red}$I_{N-2}Y_{N-1}$};

%  --------

\draw[thick, purple, very thick, dashed, rounded corners=0pt, fill=none]
    plot[smooth cycle, tension = 1] coordinates {
    ($(A)+(0,0.3)$)
    ($(B)+(0,0.2)$)
    ($(B)+(0.2,0)$)
    ($(B)+(0,-0.2)$)
    ($(A)+(0,-0.3)$)
    ($(A)+(-0.8,0)$)
    };

% --- Draw a smooth, purple dashed outline around B and the tips
\draw[thick, green!70!black, very thick, dashed, rounded corners=0pt, fill=none]
  plot [smooth cycle, tension=1] coordinates {
    ($(B)+(-0.1,-0.1)$)   % bottom-left of B
    ($(tip1)+(-0.5,0.2)$) % tip1--end of blue arrow
    % ($(tip1)+(1.1,0.4)$)
    ($(tip2)+(0,0.6)$)    % tip2--end of blue arrow
    % ($(tip2)+(1,-0.4)$)
    ($(tip3)+(0.8,0.3)$)  % tip3--end of blue arrow
    ($(B)+(0.1,-0.1)$)    % bottom-right of B
  };

\draw[thick, green!70!black, very thick, dashed, rounded corners=0pt, fill=none]
  plot [smooth cycle, tension=1] coordinates {
    ($(B)+(-0.1,-0.1)$)   % bottom-left of B
    ($(tip4)+(-0.6,-0.2)$) % tip1 with slight padding
    ($(tip5)+(0,-0.4)$)    % tip2
    ($(tip6)+(0.9,0.1)$)  % tip3
    ($(B)+(0.1,-0.1)$)    % bottom-right of B
  };

% \iffalse
% %“capsule” (rounded rectangle with big corners)
% \draw[thick, purple, dashed, rounded corners=20pt]
%   ($(box.south west)+(-0.4,-0.45)$)
%   -- ($(box.south east)+(0.4,-0.45)$)
%   -- ($(box.north east)+(0.4,0.45)$)
%   -- ($(box.north west)+(-0.4,0.45)$)
%   -- cycle;
% \fi

%---------------------------------------
% Panel B
%---------------------------------------

    \def\xshift{9}

    \node (label1) at (\xshift + -3.5, 5) {\bf\large b)};

    \node (A1) at (\xshift, 0) {\color{red} $Y_1 I_2$};
    \node (B1) at (\xshift, 2) {};
    \node (C1) at (\xshift + 2.5, 1) {};
    \node (D1) at (\xshift + 2.5, -1.3) {};

    \draw[thick, blue, -{Stealth[length=2mm]}] (A1.north) .. controls +(-0.6,0.5) and +(-0.6,-0.5) .. (B1.south);
    \draw[thick, blue, -{Stealth[length=2mm]}] (B1.south) .. controls +(0.6,-0.5) and +(0.6,0.5) .. (A1.north);

    \node at ($(A1)!0.5!(B1)$) {\color{blue} $Y_1Y_2$};

    \draw[thick, blue, -{Stealth[length=2mm]}]
    (A1.east) .. controls +(0.5,1) and +(-1,0.1) .. (C1.west);    
    \draw[thick, blue, -{Stealth[length=2mm]}]
    (C1.west) .. controls +(-0.5,-1) and +(1,-0.1) .. (A1.east);

    \node at ($(A1)!0.5!(C1)$) {\color{blue} $Z_1Z_2$};

    \draw[thick, blue, -{Stealth[length=2mm]}]
    (A1.east) .. controls +(0.5,-1) and +(-1,0.1) .. (D1.west);
    \draw[thick, blue, -{Stealth[length=2mm]}]
    (D1.west) .. controls +(-0.5,1) and +(1,-0.1) .. (A1.east);

    \node at ($(A1)!0.5!(D1)$) {\color{blue} $X_1X_2$};

    \coordinate (B1center) at (B1);
    \coordinate (B1tip1) at ($(B1center)+(135:3)$);
    \coordinate (B1tip2) at ($(B1center)+(90:3)$);
    \coordinate (B1tip3) at ($(B1center)+(45:3)$);
    
    \draw[thick, blue, dotted, -{Stealth[length=2mm]}] (B1) -- ++(135:3) node[midway, above, sloped]{\color{blue}$I_1..Y_{N-1}I_N$} node[pos=1, above, rotate=45]{\color{red}$I_1..I_{N-1}Z_N$}; 
    \draw[thick, blue, dotted, -{Stealth[length=2mm]}] (B1) -- ++(90:3) node[midway, above, sloped]{\color{blue}$I_1I_2Y_3Y_4..I_N$} node[pos=1, rotate=0, anchor=south]{\color{red} $I_1I_2I_3Y_4..I_N$};
    \draw[thick, blue, dotted, -{Stealth[length=2mm]}] (B1) -- ++(45:3) node[pos=1, rotate=-45, anchor=south] {\color{red}$Y_1Y_2I_3..I_N$} node[midway, above, sloped]{\color{blue}$I_1I_2I_3Y_4..I_N$};

    \coordinate (C1center) at (C1);
    \coordinate (C1tip1) at ($(C1center)+(45:3)$);
    \coordinate (C1tip2) at ($(C1center)+(0:3)$);
    \coordinate (C1tip3) at ($(C1center)+(-45:3)$);

    \draw[thick, blue, dotted, -{Stealth[length=2mm]}] (C1) -- ++(45:3) node[midway, above, sloped]{\color{blue}$I_1..Y_{N-1}I_N$} node[pos=1, above, rotate=-45]{\color{red}$I_1..I_{N-1}Z_N$}; 
    \draw[thick, blue, dotted, -{Stealth[length=2mm]}] (C1) -- ++(0:3) node[midway, above, sloped]{\color{blue}$I_1I_2Y_3Y_4..I_N$} node[pos=1, rotate=0, anchor=south, rotate=-90]{\color{red} $I_1I_2I_3Y_4..I_N$};
    \draw[thick, blue, dotted, -{Stealth[length=2mm]}] (C1) -- ++(-45:3) node[pos=1, rotate = 45, anchor=north] {\color{red}$Y_1Y_2I_3..I_N$} node[midway, above, sloped]{\color{blue}$I_1I_2I_3Y_4..I_N$};

    \coordinate (D1center) at (D1);
    \coordinate (D1tip1) at ($(D1center)+(-135:3)$);
    \coordinate (D1tip2) at ($(D1center)+(-90:3)$);
    \coordinate (D1tip3) at ($(D1center)+(-45:3)$);

    \draw[thick, blue, dotted, -{Stealth[length=2mm]}] (D1) -- ++(-135:3) node[midway, below, sloped]{\color{blue}$X_2X_3$} node[pos=1, rotate=-45, anchor=north]{\color{red}$I_2 Y_3$};
    \draw[thick, blue, dotted, -{Stealth[length=2mm]}] (D1) -- ++(-90:3) node[midway, above, sloped]{\color{blue}$X_4X_5$} node[pos=1, rotate=0, anchor=north]{\color{red}$I_4Y_5$};
    \draw[thick, blue, dotted, -{Stealth[length=2mm]}] (D1) -- ++(-45:3) node[midway, below, sloped]{\color{blue} $X_{N-2}X_{N-1}$} node[pos=1, rotate=45, anchor=north]{\color{red}$I_{N-2}Y_{N-1}$};

% --- Braiding centered around B1
\draw[thick, green!70!black, very thick, dashed, rounded corners=0pt, fill=none]
  plot [smooth cycle, tension=1] coordinates {
    ($(B1)+(-0.1,-0.1)$)   % bottom-left of B1
    ($(B1tip1)+(-1.3,0)$) % tip1--end of blue arrow
    ($(B1tip2)+(0,0.6)$)    % tip2--end of blue arrow
    ($(B1tip3)+(1.3,0)$)  % tip3--end of blue arrow
    ($(B1)+(0.1,-0.1)$)    % bottom-right of B1
  };

\draw[thick, green!70!black, very thick, dashed, rounded corners=0pt, fill=none]
    plot[smooth cycle, tension=1] coordinates {
    ($(C1)$)
    ($(C1tip1)+(-0.5,1)$)
    ($(C1tip2)+(0.5,0)$)
    ($(C1tip3)+(-0.2,-1)$)
    ($(C1)$)
    };

\draw[thick, green!70!black, very thick, dashed, rounded corners=0pt, fill=none]
  plot [smooth cycle, tension=1] coordinates {
    ($(D1)+(-0.1,-0.1)$)   % bottom-left of D1
    ($(D1tip1)+(-0.5,0)$) % tip1--end of blue arrow
    ($(D1tip2)+(0,-0.5)$)    % tip2--end of blue arrow
    ($(D1tip3)+(1,0)$)  % tip3--end of blue arrow
    ($(D1)+(0.1,-0.1)$)    % bottom-right of D1
  };

% --- Purple dotted line around A1, B1, C1, D1

% \draw[thick, purple, very thick, dashed, rounded corners=0pt, fill=none]
%     plot[smooth cycle, tension = 1] coordinates {
%     ($(A1.west)$)
%     ($(B1.west)$)
%     ($(B1.north)$)
%     ($(C1.north)$)
%     ($(D1.north)$)
%     ($(A1)$)
%     };
  
\end{tikzpicture}

%% file: figures/fig_quantum_circuits.tex
\begin{center}
\begin{tikzpicture}
[scale=0.75, transform shape]

% === First circuit block (top) ===
\node (label1) at (-0.5, 1.8) {\bf a)};

\begin{scope}[shift={(0,0)}]
  \node (qc1) {%
    \begin{quantikz}[row sep=0.15cm]
      & \gate[2]{ZZ} & \\
      && \\
    \end{quantikz}
  };
  \coordinate (split) at ([yshift=8pt]qc1.east);
  \draw[-{Stealth[length=0.01mm]}] (split) -- ++(1,0) coordinate (branch)
    node[midway, above]{\color{blue} $w$}
    node[midway, below]{\color{blue} $1-w$};
  \draw[-{Stealth[length=2mm]}] (branch) .. controls +(0.5,0.4) .. ++(1.2,1)
    coordinate (TopArrowEnd1)
    node[midway, above, rotate=40]{\color{blue}$XXII..$};
  \draw[-{Stealth[length=2mm]}] (branch) .. controls +(0.5,-0.4) .. ++(1.2,-1)
    coordinate (BottomArrowEnd1)
    node[midway, above, rotate=-40]{\color{blue}$II..I$};
  \node at ([xshift=1cm]TopArrowEnd1) {
    \begin{quantikz}[row sep=0.15cm]
      & \gate[2]{YY} & \\
      && \\
    \end{quantikz}
  };
  \node at ([xshift=1cm]BottomArrowEnd1) {
    \begin{quantikz}[row sep=0.15cm]
      & \gate[2]{ZZ} & \\
      && \\
    \end{quantikz}
  };
\end{scope}

% === Second circuit block ===
\begin{scope}[shift={(0,-3.8)}]
  \node (qc2) {%
    \begin{quantikz}[row sep=0.15cm]
      & \gate[2]{ZZ} & \\
      && \\
    \end{quantikz}
  };
  \coordinate (split2) at ([yshift=8pt]qc2.east);
  \draw[-{Stealth[length=0.01mm]}] (split2) -- ++(1,0) coordinate (branch2)
    node[midway, above]{\color{blue} $w$}
    node[midway, below]{\color{blue} $1-w$};
  \draw[-{Stealth[length=2mm]}] (branch2) .. controls +(0.5,0.4) .. ++(1.2,1)
    coordinate (TopArrowEnd2)
    node[midway, above, rotate=40]{\color{blue}$XXII..$};
  \draw[-{Stealth[length=2mm]}] (branch2) .. controls +(0.5,-0.4) .. ++(1.2,-1)
    coordinate (BottomArrowEnd2)
    node[midway, above, rotate=-40]{\color{blue}$II..I$};
  \node at ([xshift=1cm]TopArrowEnd2) {
    \begin{quantikz}[row sep=0.15cm]
      & \gate[2]{YY} & \\
      && \\
    \end{quantikz}
  };
  \node at ([xshift=1cm]BottomArrowEnd2) {
    \begin{quantikz}[row sep=0.15cm]
      & \gate[2]{ZZ} & \\
      && \\
    \end{quantikz}
  };
\end{scope}

% === Vertical ellipsis node ===
\node at (2,-5.5) {\Large$\vdots$};
\node at (4.3,-5.5) {\Large$\vdots$};

% === Third circuit block ===
\begin{scope}[shift={(0,-8)}]  % further down
  \node (qc3) {%
    \begin{quantikz}[row sep=0.15cm]
      & \gate[2]{ZZ} & \\
      && \\
    \end{quantikz}
  };
  \coordinate (split3) at ([yshift=8pt]qc3.east);
  \draw[-{Stealth[length=0.01mm]}] (split3) -- ++(1,0) coordinate (branch3)
    node[midway, above]{\color{blue} $w$}
    node[midway, below]{\color{blue} $1-w$};
  \draw[-{Stealth[length=2mm]}] (branch3) .. controls +(0.5,0.4) .. ++(1.2,1)
    coordinate (TopArrowEnd3)
    node[midway, above, rotate=40]{\color{blue}$XXII..$};
  \draw[-{Stealth[length=2mm]}] (branch3) .. controls +(0.5,-0.4) .. ++(1.2,-1)
    coordinate (BottomArrowEnd3)
    node[midway, above, rotate=-40]{\color{blue}$II..I$};
  \node at ([xshift=1cm]TopArrowEnd3) {
    \begin{quantikz}[row sep=0.15cm]
      & \gate[2]{YY} & \\
      && \\
    \end{quantikz}
  };
  \node at ([xshift=1cm]BottomArrowEnd3) {
    \begin{quantikz}[row sep=0.15cm]
      & \gate[2]{ZZ} & \\
      && \\
    \end{quantikz}
  };
\end{scope}

\def\xshift{7.9}

\node (label2) at (\xshift -2, 1.8) {\bf b)};
\begin{scope}[shift={(\xshift,-0.8)}]
\node (qc4) {
    \begin{quantikz}[row sep=0.15cm]
        &&\gate[8]{ZZ}& \\
        &\gate[2]{ZZ}&& \\
        &&& \\
        & \gate[2]{ZZ}&& \\
        &&& \\
        \vdots
        & \gate[2]{ZZ}&& \\
        &&& \\
        &&&
    \end{quantikz}
};
\end{scope}

\node (label3) at (\xshift -2, -5) {\bf c)};

\begin{scope}[shift={(\xshift,-7)}]
\node (qc4) {
    \begin{quantikz}[row sep=0.15cm]
        &\gate[2]{ZZ}&& \\
        &&\gate[2]{ZZ}& \\
        &\gate[2]{ZZ}&& \\
        &&\gate[2]{ZZ}& \\
        &&& 
    \end{quantikz}
};
\end{scope}
\end{tikzpicture}
\end{center}

%% file: appendix.tex
\appendix

\section{Derivation of Eq.~\eqref{channel_state_duality}}\label{app:A}
\label{app:channel_state_duality}

Following Refs.~\cite{choi1975completely,jamiolkowski1972linear,jiang2013channel},  
we start by expressing
\begin{equation}
J(\mathcal{E}) = \frac{1}{d}\sum_{i,j}\mathcal{E}(|i\rangle \langle j|) \otimes |i\rangle \langle j|\,.
\end{equation}
Taking the partial trace over the second subsystem $\mathcal{H}_2$, we obtain
\begin{equation}
\begin{split}
    \text{Tr}_2 \big[(\mathbb{I} \otimes \rho^{\top})J(\mathcal{E})\big] 
    &= \frac{1}{d}\sum_{i,j} \mathcal{E}(|i\rangle \langle j|) \langle j| \rho^{\top} |i\rangle \\
    &= \frac{1}{d}\sum_{i,j} \mathcal{E}(|i\rangle \langle j|) \langle i| \rho |j\rangle \\
    &= \frac{1}{d}\mathcal{E}\left( \sum_{i,j} \langle i | \rho | j \rangle |i\rangle \langle j| \right) \\
    &= \frac{1}{d}\mathcal{E}(\rho)\,.
\end{split}
\end{equation}

\section{Proof of Theorem~\ref{theorem}}
\label{app:bound}

We now derive upper and lower bounds on the Schatten $p$-distance between the matrices $(\mathbb{I}\otimes\rho^{\top})J(\mathcal{E}_{\text{sys}})$ and $(\mathbb{I}\otimes\rho^{\top})J(\mathcal{E}_{\text{eff}})$ for an arbitrary density operator $\rho$.

From Choi–Jamio{\l}kowski isomorphism, we have 
\begin{align}
    d\,\Tr_2\left[(\mathbb{I}\otimes\rho^{\top})J(\mathcal{E})\right]=\mathcal{E}(\rho)\,.
\end{align}
Then, we have 
\begin{align}
\norm{\mathcal{E}_{\text{sys}}(\rho)-\mathcal{E}_{\text{eff}}(\rho)}_{p}=d\,\norm{\Tr_2\left[(\mathbb{I}\otimes\rho^{\top})\Delta J\right]}_p\,.
\end{align}
where we defined 
\begin{align}
    \Delta J \equiv J(\mathcal{E}_{\text{sys}})-J(\mathcal{E}_{\text{eff}})\,.
\end{align}
From the duality property~\cite{WatrousBook18}, 
\begin{align}
    \norm{A}_p\equiv \max_{\norm{B}_q\leq 1}\Big|\Tr[B^{\dagger} A]\Big|\,,
\end{align}
where two real numberts $p$ and $q$ satisfy 
\begin{align}
    \frac{1}{p}+\frac{1}{q}=1\,,
\end{align}
we can write 
\begin{align}
\begin{split}
\norm{\Tr_2\left[(\mathbb{I}\otimes\rho^{\top})\Delta J\right]}_p& = \max_{\norm{B}_q\leq 1}\Big|\Tr\left[B^{\dagger} \Tr_2\left[(\mathbb{I}\otimes\rho^{\top})\Delta J\right]\right]\Big|\\
&=\max_{\norm{B}_q\leq 1}\Big|\Tr\left[(B^{\dagger}\otimes\mathbb{I})(\mathbb{I}\otimes\rho^{\top})\Delta J\right]\Big|\\
&=\max_{\norm{B}_q\leq 1}\Big|\Tr\left[(B^{\dagger}\otimes\rho^{\top})\Delta J\right]\Big|\\
&\leq \max_{\norm{B}_q\leq 1}\norm{B^{\dagger}\otimes\rho^{\top}}_q\norm{\Delta J}_p\\
&=\max_{\norm{B}_q\leq 1}\norm{B}_{q}\norm{\rho}_q\norm{\Delta J}_{p}\\
&=\norm{\rho}_q\norm{\Delta J}_{p}\,, 
\end{split}
\end{align}
where we used the H\"{o}lder inequality~\cite{WatrousBook18}
\begin{align}
\abs{\Tr[A^{\dagger} B]}\leq \norm{A}_{q}\norm{B}_p\,
\end{align}
and 
\begin{align}
\norm{A\otimes B}_q=\norm{A}_q\norm{B}_q\,.
\end{align}
Therefore, we have 
\begin{align}
\norm{\mathcal{E}_{\text{sys}}(\rho)-\mathcal{E}_{\text{eff}}(\rho)}_{p}\leq d\,\norm{\rho}_q\, \norm{J(\mathcal{E}_{\text{sys}})-J(\mathcal{E}_{\text{eff}})}_p\,. 
\end{align}
Since 
\begin{align}
    \norm{\rho}_q\leq 1\,,
\end{align}
we can finally obtain 
\begin{align}
0\leq \norm{\mathcal{E}_{\text{sys}}(\rho)-\mathcal{E}_{\text{eff}}(\rho)}_{p}\leq d\,\norm{J(\mathcal{E}_{\text{sys}})-J(\mathcal{E}_{\text{eff}})}_p~~(\forall \rho)\,.
\end{align}